\documentclass{article}   

\usepackage{spconf_modified,amsmath,epsfig,amssymb,amsfonts}

\usepackage[tikz]{bclogo}
\usepackage{xcolor,colortbl}
\usepackage{setspace}

\definecolor{Gray}{gray}{0.85}
\definecolor{LightCyan}{rgb}{0.88,1,1}

\newcolumntype{a}{>{\columncolor{Gray}}c}
\newcolumntype{b}{>{\columncolor{white}}c}

\title{Modulation Formats and Waveforms for the Physical Layer of 5G Wireless Networks: Who will be the heir of OFDM?}

\name{Paolo Banelli, Stefano Buzzi, Giulio Colavolpe, Andrea Modenini, Fredrik Rusek, and Alessandro Ugolini}

\address{}
\begin{document}
\maketitle 
%


\begin{abstract}
5G cellular communications promise to deliver the gigabit experience to mobile users, with a capacity increase of up to three orders of magnitude with respect to current LTE systems.  There is widespread agreement that such an ambitious goal will  be realized through a combination of innovative techniques involving different network layers. At the physical layer, the OFDM modulation format, along with its multiple-access strategy OFDMA, is not taken for granted, and several alternatives promising larger values of spectral efficiency are being considered. This paper provides a review of some modulation formats suited for 5G, enriched by a comparative analysis of their performance in a cellular environment, and by a discussion on their interactions with specific 5G ingredients. The interaction with a massive MIMO system is also discussed by employing real channel measurements.
\end{abstract}

\section*{\small INTRODUCTION}
Orthogonal frequency division multiplexing (OFDM) and orthogonal frequency division multiple access (OFDMA) are the modulation technique and the multiple access strategy adopted in long term evolution (LTE) 4G cellular network standards, respectively \cite{ghosh2010fundamentals}. 
OFDM and OFDMA  succeeded to code division multiple access (CDMA), employed in 3G  networks, for several reasons such as, to cite some,   the ease of implementation of both transmitter and receiver thanks to the use of fast Fourier transform (FFT) and inverse fast Fourier transform (IFFT) blocks, the ability to counteract multi-path distortion, the orthogonality of subcarriers which eliminates intercell interference, the possibility of adapting, the transmitted power and the modulation cardinality, and the ease of integration with multi-antenna hardware, both at the transmitter and receiver.

Nonetheless, despite such a pool of positive properties, OFDM/OFDMA are not exempt of defects, and their adoption in the forthcoming generation of wireless networks is not taken for granted. Indeed, the spectral efficiency of OFDM is limited by the need of a cyclic prefix and by its large side-lobes (which require some null guard tones at the spectrum edges),
OFDM signals may exhibit large peak-to-average-power ratio values \cite {ochiai2001distribution},
and the impossibility of having strict frequency synchronization among subcarriers makes OFDM and OFDMA  not really orthogonal techniques. In particular, synchronization is a key issue in the uplink of a cellular network wherein different mobile terminals transmit separately~\cite{Morelli2004296}, and, also, in the downlink when base station coordination is used \cite{hwang2009ofdm, irmer2011coordinated}.
For instance, with regard  to the spectral efficiency loss of side-lobes and the cyclic prefix (CP), in an LTE system operating at 10 MHz bandwidth, only 9 MHz of the band is in fact used. In addition, the loss of the CP is around 7\%, so the accumulated loss totals at 16\%.
These drawbacks, which  invalidate many of the above discussed OFDM/OFDMA advantages, form the basis of an open and intense debate on what the modulation format and multiple access strategy should be in next generation cellular networks. 5G cellular systems will feature several innovative strategies with respect to existing LTE systems, including, among others, extensive adoption of small cells, use of mm-wave communications for short-range links,  large scale antenna arrays installed on macro base stations, cloud-based radio access network, and, possibly, opportunistic exploitation of spectrum holes through a cognitive approach~\cite{haykin2005cognitive}. All these strategies will be impacted by the modulation format used at the physical layer. At the same time, 5G cellular networks will have more stringent requirements than LTE in terms of latency, energy efficiency and data rates, which again are impacted by the adopted modulation scheme.
 This paper provides a review of some of the most credited alternatives to OFDM, performs a critical mutual comparison in terms of spectral efficiency, and discusses their possible interactions with the cited technologies and requirements of 5G networks. The focus of the paper is on linear modulations and, after a quick review of OFDM, the emphasis is shifted on filterbank multicarrier, time-frequency packing, and single-carrier modulations. Particularly we will focus on spectral efficiency employing quite a general signal processing framework coupled with an information theoretic approach, which permits evaluating the practical information rate associated with a specific signal format.  The aim indeed is far away to be exhaustive with respect to all the possible implementation issues and scenarios,  still highlighting possible research directions and approaches that deserve to be further investigated. The paper is also enriched by a specific section on massive MIMO systems, and by a performance study based 
on real channel measurements from a massive MIMO testbed from Lund University.

\vspace*{-5mm}
																		
\section*{{\small SYSTEM MODEL}}\label{Sistem-Model}
For all the modulation formats  considered in this work, the complex baseband equivalent of the transmitted signal, say $x(t)$, can be expressed as
\begin{equation}
x(t) =\sqrt{PT_{\rm s}} \sum\limits_{\ell  =  - G }^{ G } {{s_\ell}(t - \ell {T_{\rm{s}}})} \; .\label{system_model}
\end{equation}
In the above equation, 
$P$ is the signal power, $T_s$ is the symbol period, $2G+1$ is the number of temporal slots spanned by each data packet, 
and the waveform $s_\ell(t)$ is the complex baseband equivalent of the waveform associated to the $\ell$-th temporal slot \cite{Matz201387, Farhang-Boroujeny201192}, and is written as
\begin{equation} \label{l-th symbol model}
s_{\ell}(t) = \frac{1}{\sqrt{N}} \sum\limits_{k = 0}^{N - 1} {{d_{k,\ell }}p(t){e^{j2\pi k\frac{{{\delta _{\rm{f}}}{\delta _{\rm{t}}}}}{{{T_{\rm{s}}}}}(t + \ell {T_{\rm{s}}})}}} \; .
\end{equation}
In \eqref{l-th symbol model}, $N$ is the number of subcarriers, $d_{k,\ell}$ is the transmitted symbol associated with the $(k,\ell)$-th resource element (that is, $k$-th subcarrier and $\ell$-th symbol interval), $p(t)$ is the underlying shaping pulse, $\delta_{\rm{t}}$ and $\delta_{\rm{f}}$ are two dimensionless constants that rule the actual time and frequency spacing among the transmitted symbols $d_{k,\ell}$. In particular, letting $T$ be a reference symbol time used for normalization and defined as  $T= \frac{T_{\rm{s}}}{\delta_{\rm{t}}}$, it is seen that symbols $d_{k,\ell}$ are spaced in time by $T_{\rm{s}} = \delta_{\rm{t}}T$ and in frequency by $\frac{\delta_{\rm{f}} \delta_{\rm{t}}}{T_{\rm{s}}}=\frac{\delta_{\rm{f}}}{T}$. Note that letting $\delta_{\rm{f}}=\delta_{\rm{t}}=1$ we obtain the usual orthogonality-preserving frequency spacing $1/T$ which holds for OFDM systems, while the dimensionless product $\delta_{\rm{f}} \delta_{\rm{t}}$ can be interpreted as a measure of how much 
symbols are packed with respect to the classical OFDM choice  \cite{Farhang-Boroujeny201192,csahin2012,Matz20111}.
Combining (\ref{system_model}) and (\ref{l-th symbol model}) we also obtain 
\begin{equation}
x(t) =\sqrt{\frac{PT_{\rm s}}{N}} \sum\limits_{k = 0}^{N - 1} {\sum\limits_{\ell = -G}^{ G } {{d_{k,\ell }}p(t - \ell {T_{\rm{s}}}){e^{j2\pi k\frac{\delta_{\rm f}{\delta_{\rm t}}}{T_{\rm s}}t}}} }\,.\label{system_model2}
\end{equation}
Note also that the shaping pulse $p(t)$ has no restrictions in its (practically finite) time duration, but it is assumed to be of unit energy, i.e., $\|p(t)\|^2=1$. Moreover, as specified later,
 variables $\{d_{k,\ell}\}$, the transmitted symbols,  are not necessarily equal to pure modulation symbols as they may include some form of signal processing, which for instance allows to consider other (staggered) lattice structures. The pure data symbols are denoted by $\{a_{k,\ell}\}$, which we assume  to be of unit average power, i.e., ${E}[|a_{k,\ell}|^2]=1$.

Now, it is easy to show that the above signal model is representative of several modulation formats.

\noindent
{\bf OFDM.} Classical OFDM systems assume $p(t)$ as a rectangular pulse of duration $T_{\rm s}=T+T_{\rm cp}$, where $T_{\rm{cp}}$ is the CP duration. Consequently,  $\delta_{\rm{t}}=1+T_{\rm cp}/T$ and  $\delta_{\rm{f}}=1$ to grant orthogonality on the useful symbol duration $T$. Note that the transmitted symbols $d_{k,\ell}$ at the edge bands can be set to zero to limit out-of-band emissions. We also recall here that OFDM is the modulation format used in the downlink of current LTE systems, whereas for the uplink an OFDM variant known as  single-carrier FDMA (SC-FDMA) is adopted, in order to limit the peak-to-average power ratio (PAPR) \cite{ghosh2010fundamentals}.

\medskip
\noindent
{\bf Filterbank multicarrier (FBMC)}.
FBMC is an OFDM-like modulation format wherein subcarriers are passed through filters that suppress signals' sidelobes, making them eventually strictly bandlimited. The transmitter and receiver may still be implemented through FFT/IFFT blocks or polyphase filter structures \cite{Farhang-Boroujeny201192, csahin2012}, and bandlimitedness may deliver larger spectral efficiency than OFDM.
The use of FBMC for 5G cellular networks is mainly endorsed for its ability (due to  signal bandlimitedness) to cope with network asynchronicity that naturally arises in the uplink and/or in the downlink with coordinated transmission \cite{wunder20125gnow}, for its greater robustness to frequency misalignments among users when compared to OFDM \cite{fusco2008sensitivity}, and for its more flexible exploitation of frequency white spaces in cognitive radio networks \cite{haykin2005cognitive,Farhang-Boroujeny201192}.
FBMC is usually either coupled with QAM or with Offset-QAM (OQAM) modulation formats. 
For FBMC-QAM, we have $\delta_{\rm{t}} \delta_{\rm{f}} \geq 1$ and the transmitted symbols $d_{k,\ell}=a_{k,\ell}$ are drawn from an ${\cal M}$-ary QAM constellation. For FBMC-OQAM, instead, 
symbols are half-spaced in time with respect to FBMC-QAM, and consequently we have $\delta_{\rm{t}} \delta_{\rm{f}} \geq 0.5$. The transmitted symbols are related to the data symbols by the relation $d_{k,\ell}=j^{k+\ell}a_{k,\ell}$, and the data symbols $a_{k,\ell}$ are real-valued $\sqrt{{\cal M}}$-ary PAM symbols.

\medskip
\noindent
{\bf Faster-than-Nyquist (FTN) / Time-frequency-packed (TFS) signaling.}
Faster-than-Nyquist (FTN) signaling, first discussed by Mazo as early as 1975 in \cite{Ma75c},  is a technique to increase the spectral efficiency of a communication system by letting $\delta_{\rm{t}}<1$, thus introducing intentional interference among data symbols at the transmitter side. FTN was for a long time studied only as a single carrier technique~\cite{LiGe03}, and over time it stood clear that FTN can exploit the excess bandwidth of the single carrier signal.
The rate gains of FTN in single carrier systems spurred a number of extensions of FTN into multicarrier setups \cite{RuAn05, Rusek2009,Han20091721,BaFeCo09b,MoRuCo14}, and the resulting modulation formats have also been named as Time-frequency-packed signaling (TFS). Let us first lay down the model for the transmitted signal of TFS. The system model we use is a generalized version of either an FBMC-OQAM model or an FBMC-QAM model. In view of (\ref{system_model2}), the TFS system has all parameters identical to its FBMC counterpart except for the product $\delta_{\rm{t}}\delta_{\rm{f}}$. When an FBMC-OQAM system underlies the TFS system, this product should satisfy $\delta_{\rm{t}}\delta_{\rm{f}}<0.5$, while for an underlying  FBMC-QAM system it satisfies $\delta_{\rm{t}}\delta_{\rm{f}}<1.$ 
More sophisticated arguments, inspired by time-frequency analysis, highlight how these communication systems are based on Weyl-Heisenberg function sets, also known as Gabor sets \cite{gabor1946,Matz201387}, and as special cases of packing data on a Grassmannian manifold \cite{Strohmer2003257}.
Arguments on data packing theory \cite{csahin2012,Strohmer2003257}, indicate that the best packing is obtained by hexagonal lattices, which provide some spectral efficiency gains with respect to rectangular, or staggered, lattices. For simplicity we will not consider this special case, although the general framework derived herein, as well as the conclusions, can be easily extended.

\medskip
\noindent
{\bf Single-carrier modulation (SCM)}
Letting $N=1$, the outlined signal model boils down to a linear single carrier modulation format. 
During recent times, multicarrier systems have been the dominant modulation format, the main reason being that optimal equalization can be efficiently carried out in the frequency domain, while optimal equalization of a single carrier system is much more involved and essentially requires the use of a Viterbi algorithm. Recently, however,  there has been regained interest in single carrier techniques due to the development of high-performance and low-complexity equalizers operating in the frequency domain  ~\cite{ohlmer2012model,Gerstacker20131,benvenuto2010single}.
In this paper, we shall consider a single carrier structure adopting a CP in order to provide an interblock interference free system and to convert the channel into a cyclic convolution, which simplifies the usage of the frequency domain equalizer, especially in time-invariant or slowly-varying channels. If the time duration of the channel impulse response is at most $T_{\mathrm{ch}}$ seconds, and the symbol time is $T_{\mathrm{s}}$, then the cyclic prefix needs to be at least $G_{\mathrm{cp}}=\lceil T_{\mathrm{ch}}/T_{\mathrm{s}} \rceil$ symbols long.
Hence, in the data block in \eqref{system_model}, only $2G+1-G_{\rm{cp}}$ symbols $d_{k,l}$ corresponds to data symbols, while the other $G_{\rm{cp}}$ represent the redundancy of the cyclic prefix.

\medskip

Table \ref{table_par} provides an overview of the values of the parameters characterizing the discussed modulations formats.

\section*{\small SPECTRAL EFFICIENCY}

When assessing the performance of a given modulation and coding system, a key figure of merit is the spectral efficiency $\rho$, defined as
$$
\rho=\frac{R_{\mathrm{c}}N\zeta_{\rm{g}}\log_2 ({\cal M})}{T_{\rm s}W_{\mathrm{tot}}}\quad\mathrm{b/s/Hz}
$$
where $R_{\mathrm{c}}$ is the rate of the employed channel code,  $W_{\mathrm{tot}}$ is the total frequency occupancy of the signal according to some measure, and $\zeta_{\rm{g}}\leq 1$ is the inefficiency due to possible guard bands in multicarrier systems, or dually, guard time in single carrier systems. We remind the reader that ${\cal M}$ denotes the cardinality of the employed modulation, and $N$ is the number of subcarriers. Note that spectral efficiency denotes here the data-rate that can be transmitted for each bandwidth unit used for transmission, {\em regardless of the underlying bit error rate (BER)}. Later on, instead, we will focus on the achievable spectral efficiency (ASE), a much more insightful performance measure,  representing the spectral efficiency which a system may attain under the constraint of arbitrarily small BER.

For OFDM, with an ${\cal M}$-ary QAM (${\cal M}$-QAM) constellation, the spectral efficiency has expression
$$
\rho_{\mathrm{OFDM}}=\frac{R_{\mathrm{c}}N\zeta_{\rm{g}}\log_2({\cal M}) }{N+G_{\mathrm{cp}}} \quad\mathrm{b/s/Hz}.
$$

Regarding FBMC, with an ${\cal M}$-QAM for FBMC-QAM, an $\sqrt{{\cal M}}$-PAM constellation for the FBMC-OQAM system, and strict equality for $\delta_{\rm{t}}\delta_{\rm{f}}$ in Table \ref{table_par}, the spectral efficiency, as $N$ grows large, in both cases becomes
\begin{equation}	
{\mathrm{FBMC}}=\log_2({\cal M})R_{\mathrm{c}}\quad\mathrm{b/s/Hz.} \label{eq:rfbmc}
\end{equation}
Thus, compared with OFDM the loss due to the CP and spectral guard bands has vanished.

Regarding spectral efficiency for SCM, instead, 
the ideal choice for $p(t)$ is a sinc pulse with double sided bandwidth $W=1/T_{\mathrm{s}}$ Hz. However, in practice this is not possible, so a smoother pulse in frequency is used. Let the bandwidth of $p(t)$ be $W_{\rm tot}=(1+\delta)W=(1+\delta)/T_{\mathrm{s}}$, where $\delta$ measures the excess bandwidth in comparison to the sinc pulse. Then, the spectral efficiency becomes
$$\rho_{\mathrm{SC}} = \frac{R_{\mathrm{c}}\zeta_{\rm g}\log_2({\cal M}) }{(1+\delta)} \quad \mathrm{b/s/Hz},$$
where $\zeta_{\rm g}=(2G+1-G_{\mathrm{cp}})/(2G+1)$ is the inefficiency due to the cyclic prefix.

Finally, regarding TFS, in a multicarrier system like FBMC, the two parameters $\delta_{\rm{t}}$ and $\delta_{\rm{f}}$ control the amount of compression of time and frequency, respectively. In the special case of $\delta_{\rm{t}}=\delta_{\rm{f}}=1$, the subcarrier spacing, in the case of a pulse shape with no roll-off,  is exactly the reciprocal of the symbol time $T_{\rm{s}}$, which means that the time-frequency occupancy per complex input symbol $d_{k,\ell}$ becomes exactly 1 Hz-s which is the smallest possible occupancy if an orthogonal set of pulses is desired. 
For TFS-QAM with time and frequency packing activated, i.e., $\delta_{\rm{t}}\delta_{\rm{f}}<1$, an ${\cal M}$-QAM constellation, and a rate $R_{\mathrm{c}}$ code, the spectral efficiency becomes
\begin{equation}
\rho_{\mathrm{TFS-QAM}}=\frac{R_{\mathrm{c}}\log_2({\cal M})}{\delta_{\rm{t}}\delta_{\rm{f}}}\quad\mathrm{b/s/Hz.} \label{eq:tfsqam}
\end{equation}
For TFS-OQAM with an $\sqrt{{\cal M}}$-PAM constellation, the spectral efficiency becomes
\begin{equation}
\rho_{\mathrm{TFS-OQAM}}=\frac{R_{\mathrm{c}}\log_2({\cal M})}{2\delta_{\rm{t}}\delta_{\rm{f}}}\quad\mathrm{b/s/Hz.} \label{eq:tfsoqam}
\end{equation}
Thus, a spectral efficiency gain proportional to $1/\delta_{\rm{t}}\delta_{\rm{f}}$ is achieved compared with an FBMC system, at the cost of increased interference among the symbols $\{d_{k,\ell}\}$. Note that, by setting the limit values of $\delta_t\delta_f$ in the two equations  
($\delta_t\delta_f=1$ and $\delta_t\delta_f=0.5$ respectively), \eqref{eq:tfsqam} and \eqref{eq:tfsoqam} collapse into \eqref{eq:rfbmc}. 
In small cells, where the SNR can be very high, the current trend is to employ high-order constellations, such as 256-QAM or 1024-QAM. This is in sharp contrast to TFS, which maintains a small constellation size, such as quaternary PSK (QPSK) or 16-QAM, but increases the degree of time-frequency compression in order to achieve higher spectral efficiencies.

\section*{\small DISCUSSION AND LITERATURE OVERVIEW}

As already commented, OFDM is a multicarrier modulation format wherein the use of a CP and a proper spacing among subcarriers ensure orthogonality of the waveforms modulated by different data symbols. In general, the amount of interference among adjacent (both in time and frequency) data symbols is ruled by the sampling 
on the time-frequency plane of the ambiguity function associated to the prototype pulse shape $p(t)$, expressed by~\cite{Farhang-Boroujeny201192,Matz201387,csahin2012}
\begin{equation}
A_p(\tau,\nu)=\int\limits_t {p(t){p^*}(t -\tau){e^{-j2\pi\nu t}}{\rm d}t}\,.
\end{equation}
Thus, in order to minimize the interference from adjacent symbols in time (intersymbol interference, ISI), and in frequency (interchannel interference, ICI), several research efforts have been dedicated to designing pulse shapes with good ambiguity functions, i.e., according to an orthogonal design in both the domains, as expressed by $A_p(\ell \delta_{\rm t}T,k \delta_{\rm f}/T)=\delta[k]\delta[\ell]$, where $\delta[i]$ is the Kronecker delta function.
An excellent overview in this respect is provided in \cite{csahin2012}. However, double orthogonal designs exist only when $\delta_{\rm t}\delta_{\rm f} > 1$. Furthermore, multipath channels could destroy orthogonality, and mismatched filtering may be preferable in this case \cite{Farhang-Boroujeny201192}. Anyway, also with mismatched filtering, double-domain orthogonality can be granted (with some SNR penalty in AWGN) only when \cite{csahin2012,Farhang-Boroujeny201192} $\delta_{\rm t}\delta_{\rm f} > 1$.
For instance, OFDM preserves orthogonality in frequency-selective (multi-path) channels by adding a guard time between successive symbols, by means of a CP or  zero padding (ZP) \cite{Wang200029,Muquet20022136} which leads to $\delta_{\rm t}\delta_{\rm f}=1+T_{\rm{cp}}/T > 1$.
Furthermore, constraining the symbols to be real (or imaginary), i.e., using PAMs rather than QAMs, orthogonality can be granted also when $\delta_{\rm t}\delta_{\rm f}=1/2$, as already noticed in the first studies about multicarrier systems, \cite{chang1966,saltzberg1967}, and successively called Offset-QAM-based OFDM (OFDM-OQAM)  \cite{hirosaki1981}, which realizes a rectangular lattice staggering in the time-frequency plane.

Regarding OFDM, orthogonality is lost in the presence of frequency synchronization errors or phase noise, which cause non negligible performance loss to OFDM(A) systems \cite{Rugini20052279,Cai20032047,Tomba1998580,Morelli2004296}.
Furthermore, orthogonality is also lost (and performance significantly degrades) if any carrier frequency offset (CFO) is present \cite{Rugini20052279} or the multipath channel  is significantly time-varying (doubly-selective) within the symbol period $T_{\rm s}$. In this case, interference cancellation/mitigation techniques should be considered also for the orthogonally designed OFDM systems \cite{Schniter20041002,Mostofi2005765,Rugini20061,Rugini2011285}. This fact is one of the main motivations for recent research efforts on filterbank multicarrier (FBMC) schemes that, exploiting similar approaches to combat ISI and ICI, by proper pulse shape designs may combat the sensitivity to CFO and doubly-selective channels, still preserving spectral efficiency with $\delta_{\rm t}\delta_{\rm f} =1$ \cite{Fusco20092705}. Actually, the same philosophy can be used also when $\delta_{\rm t}\delta_{\rm f} < 1$~\cite{Han20091721}, e.g., with the generalization of 
FBMC according to an FTN principle \cite{Ma75c,RuAn05,Rusek2009,BaFeCo09b}.
The idea is that relaxing the orthogonality constraints \cite{Kozek19981579}, the pulse shape design has higher degrees of freedom to reduce ISI and ICI sensitiveness under doubly-selective channels or CFO effects.

Turning back to the issue of spectral efficiency maximization, we note that  ideally a sinc pulse should be used in single carrier FTN. In practice a smoother spectrum with roll-off $\delta$ is instead employed. In an AWGN channel and without TFS, the adoption of these pulses would result in a loss of a factor $1/(1+\delta)$ from the Shannon limit in terms of spectral efficiency. However, it can be proved that with FTN the maximum overall spectral efficiency, even when $\delta>0$, tends to the Shannon limit. Hence, with FTN the excess bandwidth is not imposing any loss at high SNR~\cite{RuAn06}.

Unfortunately, the impressive rate gains of single carrier TFS do not in general carry over to FBMC systems. The reason is that with FBMC, the excess bandwidth is typically smaller and limited to the last subcarriers at the band edge. Yet, there are reasons that motivate why TFS may still be attractive in multicarrier systems, which we discuss next.

First, one of the constraints in the design of a classical FBMC system is that the time and frequency translated pulses should form an orthonormal basis so that the data symbols can be demodulated independently in an AWGN channel. However, in all channels, save for the special case of an AWGN channel, orthogonality of the pulses is anyway lost at the receiver. This requires some form of an equalizer structure at the receiver side, and such equalizer can just as well be designed so that it also equalizes the self-induced interference. Still, the constraint of orthogonality puts heavy restrictions on the FBMC design and, by relaxing it, additional degrees of freedom in the pulse design are made available at the cost of a controlled amount of interference.

In cases where the allocated bandwidth to one user is small, the amount of excess bandwidth at the band edge can still be relatively large. In such cases, TFS can beneficially exploit the side lobes to increase the spectral efficiency. Take the LTE system as an illustrative example: in LTE's 1.4 MHz downlink mode, only 1.08 MHz of the band is used for transmission. The remaining 0.32 MHz is a guard band, and does not contribute to the data rate. With TFS, the guard band starts contributing to the data rate, giving an improved spectral efficiency.

Lastly, TFS offers a flexible method to adapt the spectral efficiency through varying the two compression parameters $\delta_{\rm{t}}$ and $\delta_{\rm{f}}$. With FBMC, the data rate can only be adapted in discrete steps by changing the constellation size and the coding rate. With TFS, a much finer granularity is achieved. As an extra bonus, TFS may also reduce the number of error correcting codes as it can maintain the same code rate but adapt the spectral efficiency by controlling the parameters $\delta_{\rm{t}}$ and $\delta_{\rm{f}}$. Moreover, TFS creates a shaping effect of the input constellation, so that an SNR gain is typically achieved over standard QAM-type constellations.

\section*{\small DISCRETE-TIME MODEL}\label{Vector-Matrix model}
In what follows we outline a discrete-time model for the considered modulations, which can be obtained by sampling the general waveform in \eqref{l-th symbol model}; moreover, we give an expression for the discrete-time received signal after it has passed through a (possibly) time varying channel with impulse response $h_{\rm c}(t,\tau)$.

Thus, by employing a sampling frequency $F_{\rm c}=1/T_{\rm c}=\frac{N_{\rm s}}{T_{\rm s}}$, with $N_{\rm s}$ an integer representing the oversampling factor, the discrete-time signal $s_{\ell}[n]={s}_{\ell}(nT_{\rm c})$ associated to the $\ell$-th symbol is expressed by
\begin{equation}\label{l-th symbol short}
\begin{array}{rl}
s_{\ell}[n] =& \frac{1}{\sqrt{N}} \sum\limits_{k=0}^{N-1}{{{{\tilde{d}}}_{k,\ell}}\underbrace{p[n]{{e}^{j2\pi k\frac{{\delta_{\rm f}}{\delta_{\rm t}}T_{\rm c}}{T_{\rm s}}n}}}_{{{p}_{k}}[n]}}
\\
=& \frac{p[n]}{\sqrt{N}} \underbrace{\sum\limits_{k=0}^{N-1}{{{{\tilde{d}}}_{k,\ell}}{{e}^{j2\pi k\frac{{\delta_{\rm f}}{\delta_{\rm t}}T_{\rm c}}{T_{\rm s}}n}}}}_{d_{\ell}[n]},
\end{array}
\end{equation}
where $p[n]=p(nT_{\rm c})$ is the discrete-time pulse shape during the time support $[0,(Q-1)T_{\rm c}]$.
Note that, when ${\delta_{\rm f}}{\delta_{\rm t}}\ne 1$, this corresponds to transmitting phase-rotated symbols (differently for each subcarrier and symbol period), as expressed by $\tilde{d}_{k,\ell}={{d}_{k,\ell}}{{e}^{j2\pi {\delta_{\rm f}}{\delta_{\rm t}}k\ell}}$.
The first equality in \eqref{l-th symbol short} highlights that the signal is obtained by multiplexing the data by a bank of filters $p_{k}[n]$, while the second shows that the signal is also a time-domain windowing $p[n]$, independent of $\ell$, of a multicarrier (OFDM-like) signal $d_{\ell}[n]$.
Collecting the transmitted samples in a $Q \times 1$ vector $\mathbf{s}_{\ell}=\left[ s_{\ell}\left(0 \right),\ldots , s_{\ell}\left( nT_{\rm c} \right),\ldots, s_{\ell}\left( \left( Q-1 \right)T_{\rm c}\right) \right]^{T}$  the two equalities suggest equivalent block-matrix representations, as expressed by
\begin{equation}
\begin{array}{rl}\label{symbol-vector-s_l}
  \mathbf{s}_{\ell} =& \sum\limits_{k=0}^{N-1}\tilde{d}_{k,\ell} \mathbf{p}_{\rm{t}_{k}}
  =\underbrace{\left[ \mathbf{p}_{\rm{t}_{0}},\ldots ,\mathbf{p}_{\rm{t}_{k}},\ldots ,\mathbf{p}_{\rm{t}_{N-1}} \right]}_{\mathbf{P}_{\rm{t}}} \mathbf{\tilde{d}}_{\ell}
  \\
   =& \mathbf{F}_{Q}^{H}\underbrace{\left[ \mathbf{p}_{\rm{f}_{0}},\ldots ,\mathbf{p}_{\rm{f}_{k}},\ldots ,\mathbf{p}_{\rm{f}_{N-1}} \right]}_{\mathbf{P}_{\rm{f}}} \mathbf{\tilde{d}}_{\ell}
  \\
  =& \rm{diag}\left( \mathbf{p}_{\rm{t}_{0}} \right) \underbrace{\left[ \mathbf{\tilde{f}}_{0},\ldots ,\mathbf{\tilde{f}}_{k},\ldots ,\mathbf{\tilde{f}}_{N-1} \right]}_{\mathbf{\tilde{F}}^{H}}\mathbf{\tilde{d}}_{\ell} \\
\end{array}
\end{equation}
where $\mathbf{\tilde{d}}_{\ell}$
represents the $N \times 1$ transmitted data with $\left[\mathbf{\tilde{d}}_{\ell}\right]_{k}={\tilde{d}}_{k,\ell}$, ${{\mathbf{p}}_{{{\rm{t}}_{k}}}}$
 is the $Q \times 1$ $k$-th discrete-time pulse shape with $\left[\mathbf{p}_{\rm{t}_{k}}\right]_n= p_{k}[n]$, ${{\mathbf{F}}_{Q}}$ is a $Q\times Q$ unitary DFT matrix with $\left[ \mathbf{F} \right]_{k+1,n+1}=\frac{1}{\sqrt{Q}} e^{-j\frac{2\pi}{Q}kn}$, $\mathbf{p}_{\rm{f}_{k}}=\mathbf{F}_{Q}\mathbf{p}_{\rm{t}_{k}}$
represents the $k$-th pulse shape in the discrete frequency-domain, and $\mathbf{\tilde{F}}$ is a $N \times Q$ pseudo-DFT matrix with ${\left[ \mathbf{\tilde{F}} \right]}_{k+1,n+1}=\frac{1}{\sqrt{N}}e^{-j2\pi\delta_{\rm{f}}\delta_{\rm{t}}\frac{T_{\rm c}}{T_{\rm s}}kn}$, whose row-vectors $\mathbf{\tilde{f}}_{k}^{H}$ represent the modulation frequencies.
Note that, the signal vector $\mathbf{s_{\ell}}$ is obtained by a prototype pulse-shaping filter $p(t)$ that spans $\left\lceil Q/N_{\rm s} \right\rceil$ consecutive blocks, which are transmitted every $T_{\rm s}=N_{\rm s} T_{\rm c}$ seconds. Thus, each symbol would generate ISI to the adjacent ones, unless it is designed according to an orthogonal paradigm, e.g., by a Nyquist principle.

Observing \eqref{l-th symbol short}, and that a time-domain multiplication induces a circular convolution in the DFT domain, if the signal parameters are chosen such that the DFT frequency bins are aligned with the modulation frequencies, i.e., if
$$
Q=\frac{1}{{{\delta }_{\text{f}}}{{\delta }_{\text{t}}}}MN_{\text{s}} \qquad Q,M,N_{\text{s}}, \in \mathbb{N}
$$
the matrix $\mathbf{\tilde{F}}_{{}}^{H}$ is obtained collecting the equi-spaced (by $M$) rows of the $Q\times Q$ IDFT matrix ${\mathbf{F}_{Q}^H}$.

We define $\mathbf{Z}^n$ ($\mathbf{Z}^{-n}$) as the Toeplitz matrix  with all zeroes, but ones in the $n$-th sub-diagonal (super-diagonal). The transmitted vector during the $\ell$-th symbol period is thus expressed as
\begin{equation}\label{l-th transmitted vector}
\mathbf{x}_{\ell}=\sum\limits_{m}{{{\mathbf{Z}}^{mN_{\rm s}}}{{\mathbf{s}}_{\ell+m}}}=
\sum\limits_{m} \mathbf{Z}^{mN_{\rm{s}}}\mathbf{F}_{Q}^{H} \mathbf{P}_{\rm{f}} \mathbf{\tilde{d}}_{\ell+m}.	
\end{equation}
Denoting, as already specified, by $h_{\rm c}(t,\tau)$ the (possibly) time-varying channel, $h_{i,j}^{\rm (c)}=h_{\rm c}(iT_{\rm c},jT_{\rm c})$ is the discrete-time counterpart, and $\mathbf{H}_{\rm{c},\ell}^{\rm (t)}$  the $Q \times Q$  channel matrix that processes the signal transmitted at the $\ell$-th symbol period, with $\left[\mathbf{H}_{\rm{c},\ell}^{\rm (t)}\right]_{i+1,j+1}=h_{{\ell}N_{\rm{s}}+i,i-j}^{\rm (c)}$.
In order to recover the data ${{\mathbf{\tilde{d}}}_{\ell}}$ transmitted with the $\ell$-th data-block, it is necessary to observe the channel output for (at least) $ QT_{\rm c}$ seconds, and the associated received vector is expressed by
\begin{equation}\label{l-th time_received_vector}
\begin{array}{rl}
\mathbf{y}_{\ell} =& \mathbf{H}_{\rm{c},\ell}^{\rm{(t)}}{{\mathbf{x}}_{\ell}}+\mathbf{w}_{\ell} =\sum\limits_{m} \mathbf{H}_{m,\ell} \mathbf{d}_{\ell+m} +\mathbf{w}_{\ell}
\\
=& \mathbf{H}_{\rm{tot},\ell}^{\rm{(t)}}\mathbf{d}_{\ell}^{\rm{(long)}}+{{\mathbf{w}}_{\ell}},
\end{array}
\end{equation}
where $\mathbf{d}_{\ell}^{\rm{(long)}} = \left[ \ldots ,\mathbf{\tilde{d}}_{\ell-1}^T,\mathbf{\tilde{d}}_{\ell}^T,\mathbf{\tilde{d}}_{\ell+1}^T, \ldots \right]^T$ is the vector containing both the data of interest and the interference, and $\mathbf{w}_\ell$ represents the noise at the receiver. 
In the light of the multicarrier modulation format, the observation model can also be conveniently expressed in the frequency domain by projecting $\mathbf{y}_{\ell}$ on the same DFT grid of size $Q$, obtaining
\begin{equation}\label{l-th frequency_received_vector} 
\begin{array}{rl}
\mathbf{y}_{\ell}^{\rm{(f)}}=&\mathbf{{F}}_{Q}\mathbf{y}_{\ell}=\mathbf{H}_{\rm{tot}}^{\rm{(f)}}\mathbf{d}_{\ell}^{\rm{(long)}}+\mathbf{F}_{Q}\mathbf{w}_{\ell}
\\
=&\mathbf{H}_{\rm{c}}^{\rm{(f)}}\sum\limits_{m}\mathbf{C}_{m} \mathbf{P}_{\rm{f}} \mathbf{d}_{\ell+m} +\mathbf{w}_{\ell}^{\rm{(f)}}
\end{array}
\end{equation}
where  $\mathbf{H}_{\rm{tot}}^{\rm{(f)}}=\mathbf{{F}}_{Q} \mathbf{H}_{\rm{tot}}^{\rm{(t)}}$ is the total observation matrix in the frequency domain, $\mathbf{H}_{\rm{c}}^{\rm{(f)}}=\mathbf{{F}}_{Q}\mathbf{H}_{\rm{c}}^{\rm{(t)}}\mathbf{F}_{Q}^{H}$ is the frequency-domain channel matrix, ${\mathbf{C}}_{m}=\mathbf{F}_{Q} \mathbf{Z}^{mN}\mathbf{F}_{Q}^{H}$ is a full (diagonally dominant) matrix that modifies the pulse shaping matrix $\mathbf{P}_{\rm{f}}$ (note that $\mathbf{C}_{0}=\mathbf{I}_{Q}$), and we omit in the following the dependence on $\ell$ of the channel matrices for notation compactness. 

It is worth noting that the observation models in \eqref{l-th time_received_vector}-\eqref{l-th frequency_received_vector} share high similarities with the equalization of OFDM signals in doubly-selective channels, and several linear and non-linear data receiver structures may be borrowed, possibly including (data-aided) ISI and ICI cancellation \cite{Mostofi2005765,Rugini20061,Rugini2011285,Matz20111}, as well as joint iterative (turbo) equalization and decoding, \cite{Schniter20041002,Fang2008}.
Here for simplicity, we assume separate data equalization from decoding, and we only consider linear approaches such as matched-filtering (MF), least squares (LS), and linear MMSE.
Focusing on the frequency-domain observation model, the (soft) estimates of the transmitted data is generally expressed by
$$
\mathbf{\hat{d}}_{\ell} = \rm{diag}\left({\boldsymbol{\epsilon}_{\ell}}\right)\mathbf{G} \mathbf{y}_{\ell}^{\rm{(f)}},
$$
where $[\boldsymbol{\epsilon}_{\ell}]_k=e^{-j2\pi \delta_{\rm{t}}\delta_{\rm{f}} k \ell }$ compensates the phase-rotation when $\delta_{\rm{t}}\delta_{\rm{f}}\ne 1$, $\mathbf{G}$ contains the central rows of the full equalization matrices $\mathbf{G}_{\rm{MF}}={\mathbf{H}_{\rm{tot}}^{\rm{(f)}}}^{H}$, $\mathbf{G}_{\rm{LS}}={\mathbf{H}_{\rm{tot}}^{\rm{(f)}}}^{\dag}$, or $\mathbf{G}_{\rm{MMSE}}=
{\mathbf{H}_{\rm{tot}}^{\rm{(f)}}}^{H}\left(  \mathbf{H}_{\rm{tot}}^{\rm{(f)}}{\mathbf{H}_{\rm{tot}}^{\rm{(f)}}}^{H} + \sigma_{n}^2\mathbf{I}_Q \right)^{-1} $ \cite{kay1998}.
\medskip

\noindent
{\bf Extensions to other modulation formats and discussion.}
As anticipated in the system model section, the data vector may contain some signal processing of the real information vector $\mathbf{a}_{\ell}$, which in the linear case, can be captured by a precoding matrix, e.g., $\mathbf{d}_{\ell}=\boldsymbol{\Theta}{\mathbf{a}_{\ell}}$, and the equalization/detection strategy modified accordingly.
A sort of precoding $\mathbf{\Theta_{\rm{p}}}$ on non-finite alphabets, actually a prefiltering, may be also applied to each vector $\mathbf{s}_{\ell}$. This way, by proper definition of the pre-filtering/pre-coding matrices $\mathbf{\Theta}_{\rm{p}}$, $\mathbf{\Theta}$, also Generalized-FDM (GFDM) \cite{Fettweis2009}, as well as Universal-FDM (UFDM) \cite{5GNow-D3.1-2013}, can be casted in this framework. Note that, classical FBMC in \eqref{l-th symbol short}-\eqref{symbol-vector-s_l} performs a pre-filtering in the time domain by a diagonal matrix, and consequently a circular precoding on the data $\mathbf{d}_{\ell}$. However, a different structure can be imposed to the pre-filtering matrix, such as to be full in the time-domain and block-diagonal in the frequency domain, jointly performing spectrum shaping on a block of subcarriers, rather than separately on each one, as proposed for UFDM in the ongoing research project 5GNow \cite{5GNow-D3.1-2013}, and somehow reminescent of sub-blocks precoding in \cite{Wang200029}.
Adaptation to multi-antenna systems is also straightforward by collecting data and observation vectors at each antenna, leading to an observation matrix whose size increases proportionally to the number of antennas.
Obviously, the overall complexity, as well as the amount of (inter-antenna) interference will increase, making more challenging the use of MIMO and STC (Alamouti) systems, which heavily rely on the orthogonality (e.g., absence of ISI/ICI) offered by OFDM in frequency-selective channels. This has probably been one of the strongest objections for employment of FBMC-like systems so far. However, several researchers have already proposed algorithms to deal with these problems, within the great effort of the PHYDIAS project \cite{PHYDIAS-D3.1-2008} to establish and promote FBMC-based wireless communications (see \cite{Ihalainen20112070,Farhang-Boroujeny201192} and references therein).
Generally, observing that also OFDM faces almost the same problem in doubly-selective channels, where it suffers only ICI, channel estimation algorithms, receiver structures, and overall system design can take inspiration by the abundant literature on this subject \cite{Stamoulis20022451,Barhumi20042055}. For instance, receiver time-domain windowing is effective in this sense to boost the signal-to-noise plus interference ratio (SINR), as proposed in \cite{Schniter20041002,Rugini20061} for pure OFDM. Actually, transmitter and receiver time-domain windowing have been jointly optimized in \cite{Das20075782} in multicarrier communications without CP. Recently, maximum SINR approaches for MIMO-FBMC have been investigated in \cite{Caus2014,Caus20126519}, showing negligible performance degradation with respect to OFDM and significant performance gain with respect to the first attempts to MIMO-FBMC \cite{PHYDIAS-D3.1-2008,Farhang-Boroujeny201192}.

 \section*{\small PERFORMANCE ASSESSMENT} \label{discussion}

The modulation formats here discussed will now be compared when used in a typical cellular environment. In particular, we considered the Extended Typical Urban (ETU) channel defined in \cite{LTE10}. This is an example of time- and frequency-selective channel whose continuous-time impulse response can be modeled as
\begin{equation}\label{eq:cont_t_mod}
h_{\rm{c}}(t,\tau)=\sum_{k} c_{k}(t) \delta(\tau-\tau_k)
\end{equation}
where fading coefficients $c_{k}(t)$ and continuous-time delays $\tau_k$ are typical of each analyzed scenario, and $\delta(\tau)$ is the Dirac delta. From the model (\ref{eq:cont_t_mod}), the following discrete-time model has been adopted
$$
h_{i,j}^{\rm{(c)}}=\sum_{k} c_{k,i} \delta[j-j_k]
$$
where coefficients $c_{k,i}$ have been generated according to~\cite{ZhCh03}, and the discrete-time delays $j_k$ have been chosen as an approximation of continuous-time delays in (\ref{eq:cont_t_mod}) to their closest integer-multiple of $T_{\rm c}$.

As mentioned, a key figure of merit is represented by the ASE. Indeed, by properly tailoring the channel code to the considered modulation and the actual channel characteristics, the ASE performance can be closely approached. On the contrary, a comparison based on the bit or packet error rate (PER) performance for a given code does not result to be fair since the code must be specifically tailored to the considered modulation format. Generally speaking, and assuming a quasi-static channel, a code designed for the additive white Gaussian noise channel is expected to work well jointly with OFDM or FBMC with offset (or other orthogonal signaling formats). On the contrary, this kind of code will exhibit a significant performance degradation when used with modulation formats which explicitly introduce ISI and/or ICI, as TFS or FBMC without offset. By considering the above mentioned channel we are also, implicitly, assessing the robustness of the considered modulation schemes against multipath.

We will assume perfect channel state information at the receiver. Thus, our analysis does not take into account the degradation due to an imperfect channel estimation and the different losses, in terms of spectral efficiency, due to possible different requirements in terms of training or pilot sequences inserted for an accurate channel estimation.

The ASE results will be reported as a function of the ratio between the signal power $P$ and the noise power $P_{\rm n}$ computed on a reference bandwidth of 1.92 MHz. 
The spectra of the considered pulses $p(t)$ are reported in Fig. \ref{fig:pulses}. In the figure, we see the sinc pulse adopted by OFDM, a pulse with RRC spectrum and excess of bandwidth of 10\% (filter length $Q=1280$, $M=10$) adopted for FBMC-OQAM and single carrier modulations, and the pulse proposed in the PHYDIAS project \cite{PHYDIAS-D5.1-2008} for FBMC (filter length $Q=640$, $M=5$), where its improved frequency selectivity has been
accomplished by using a longer and spectrally well-shaped prototype filter.

In all cases, for OFDM, we will set the losses due to the cyclic prefix and to the insertion of a number of guard tones compliant with the LTE standard, i.e., 16\% in terms of ASE. Moreover, the transmitted symbols, for all the waveforms, will be affected by a random error vector magnitude (EVM) of 4\%, with the aim of modeling various imperfections in the implementation (such as carrier leakage, phase noise, etc.). The fractional MMSE equalizer  $\mathbf{G}_{\rm MMSE}$ is adopted at the receiver for all schemes with the exception of FBMC-OQAM, for which we used a matched filter followed by an MMSE equalizer. Indeed, the  required length of the filter, for orthoghonality in FBMC-OQAM, makes the receiver complexity too high to use a fractional MMSE. 

Figs. \ref{fig:ase_etu_fd0} and \ref{fig:ase_etu_fd30k} show the ASE performance for the two extreme scenarios of very low- and high-mobility,  characterized by Doppler frequencies $f_\mathrm{d}=0$ and $f_\mathrm{d}=30$kHz on the ETU channel. We consider
QAM with high cardinality $\mathcal{M}=64$.
The figures compare OFDM with FBMC, when $N=128$ carriers are spaced by 15 kHz. For comparison, we also show the ASE curve of a single carrier with CP system with the same bandwidth 1.92 MHz.
We see that, for the low-mobility case, OFDM, FBMC-QAM and FBMC-OQAM have similar performance: FBMC-OQAM achieves a higher spectral efficiency w.r.t. other modulations for low and medium $P/P_{\rm n}$, but at high $P/P_{\rm n}$ values it is outperformed since it has a limited complexity receiver.
Instead, performance of SCM is quite limited, since its waveform is strongly affected by the frequency selectivity of the channel, and ASE is limited by the CP loss.
In case of high-mobility, we see that FBMC-OQAM performance collapses, since its orthogonality is completely destroyed. Instead, FBMC-QAM is more resistant to Doppler and it also gains w.r.t. OFDM.

Since high-order constellations are more sensitive to the impact of the interference present when non-orthogonal signaling is adopted, we also studied the same scenario when the modulation has cardinality $\mathcal{M}=4$. We can see from Fig. \ref{fig:ase_etu_fd30k_4QAM} that, for this scenario, FBMC outperforms all other modulation formats.
We also point out that for all the considered channels, FBMC gains can be even higher by means of a properly designed pulse~\cite{csahin2012,Han20091721}.

We now consider the same scenario when TFS is adopted. Fig.~\ref{fig:ase_etu_tfp_f} shows the performance of TFS when $\mathcal{M}=4$. Different spacing values $\delta_{\rm t}$ and $\delta_{\rm f}$ have been considered and, to have a wider insight on the possible benefits of this technique, we report the highest ASE achievable when packing in the time domain only ($\delta_{\rm t}=0.90$), in the frequency domain only ($\delta_{\rm f}=0.95$), and in both domains ($\delta_{\rm t}=0.90$ and $\delta_{\rm f}=0.95$) is adopted. We can see that TFS gains with respect to FBMC are limited, and only at high $P/P_{\rm n}$. This is, in some way, expected:  further gains could be obtained with more complex receivers (techniques of advanced trellis processing~\cite{Ma75c,PiMoCoAl13}) but, on the other hand, it could be difficult to find substantial gains, since FBMC is already a sort of time-frequency packing. Our own feeling is that these are first results and more research on this topic is required.


As already discussed, the ASE can be approached in practice with proper modulation and coding  formats. Fig.~\ref{fig:ber} shows the BER of OFDM and FBMC-QAM for the scenario of Fig.~\ref{fig:ase_etu_fd30k}.  The adopted codes are low-density parity-check codes with rate 1/2 and blocklength 64,800 bits. 
In all cases, a maximum of 50 decoder iterations were performed. We can notice that performance is in accordance with the ASE results. 
We point out that the loss from the theoretical limit is twofold: first, the adopted code has finite length. Second, it is not designed for the considered channel: the use of codes properly designed for this kind of channels can considerably reduce the loss.

Summarizing, as already anticipated in the introduction, we are far away to establish which should be the preferred system, because results highly depend on the specific scenario.  Thus, extensive work still has to be done in order to identify optimal design strategies, which include setting the optimal right number of carriers (possibly different for different signal waveforms, especially in the presence or the absence of CP), the optimal pulse-shaper design, which may strongly depend on the information available at the transmitter about the  channel maximum delay spread, maximum Doppler spread, and amplitude statistics, and the optimal length of the CP, as suggested for example in \cite{whatwillbe}, wherein an OFDM system with tunable length of the cyclic prefix is proposed.

\section*{\small SINGLE CARRIER FTN MULTI-USER MODULATION WITH MASSIVE MIMO}\label{Massive MIMO}
5G Macro base stations will certainly be equipped with large scale antenna arrays, a technology also known as massive MIMO \cite{rusek2013scaling,hoydis2013massive}. Using a large number of antennas will help to boost the network throughput, since accurate beamforming will permit serving several users in the same cell and on the same bandwidth, and to stabilize the propagation channel by reducing channel outages by virtue of diversity. The joint design of interference coordination schemes and modulation formats for massive MIMO systems is a topic that will certainly gain momentum in the coming years. 

Let us discuss the uplink between $U$ single antenna users and a base station equipped with $N_{\rm BS}$ antennas. The impulse response, assumed time-invariant for simplicity, between the $u$-th user and the $n$-th base station antenna is denoted by $h_{u,n}(t)$, and we assume these to be perfectly known at the base station side. Each user transmits a, CP-free, single carrier FTN signal according to $$x_{u}(t)=\sqrt{\frac{PT_{\rm s}}{N_{\rm BS}}} \sum_{\ell}d_{u,\ell}p(t-\ell T_{\rm s}).$$ The array gain is here harvested as a power saving at the user side and not as increased signal strength at the base station side.
The received signal at the $n$-th antenna becomes
$$y_{n}(t)=\sum_{u=1}^U \sqrt{\frac{PT_{\rm s}}{N_{\rm BS}}} \sum_{\ell}d_{u,\ell}z_{u,n}(t-\ell T_{\rm s}) + n(t),$$
where $z_{u,n}(t)$ is the received pulse from the $u$-th user at the $n$-th antenna, i.e., $z_{u,n}(t)=p(t) \star h_{u,n}(t),$ where ``$\star $'' denotes convolution. To keep complexity low, we consider only single-user detection and construct a discrete-time sequence ${\bf y}_u = \{y_{u,\ell}\}$ for the detection of user $u$ according to
$$y_{u,\ell} = \sum_{n=1}^{N_{\rm BS}} y_n(t) \star z_{u,n}^{\ast} (-t) \bigg|_{t=\ell T_{\rm s}}.$$

In the case of no FTN, the receiver model is
$y_{u,\ell}=\gamma_{u} d_{u,\ell}+\eta_{u,\ell}$
where $\gamma_{u}$ is a measure of signal strength for the $u$-th user and $\eta_{u,\ell}$ collects noise, intersymbol interference, and interuser interference. Under the assumptions that all channel impulse responses are independent and that rich scattering is present, the effect of letting $N_{\rm BS}$ grow is that the impact of intersymbol and interuser interference becomes less and less;  asymptotically they both vanish. In such favorable propagation environments, there is no need for any multicarrier system to mitigate multi-path as one-tap equalizers can be used, and several users can be spatially multiplexed, which increases spectral efficiency. 

While a single carrier system has lower PAPR compared with FBMC systems, there is a reduction of spectral efficiency since pulses with excess bandwidth of an amount $\delta$ must be used. 
To reduce the loss of the excess bandwidth, we make use of FTN. Also in this case it holds that ICI vanishes as $N_{\rm BS}$ grows, but it is no longer true that ISI vanishes. Therefore, we must model the sequence ${\bf y}_u$ as
$${\bf y }_u = {\bf g}_u \star {\bf d}_u +{\bf \eta}_u,$$
where $ {\bf g}_u$ is the effective impulse response for user $u$ and ${\bf \eta}_u$ collects interuser interference and noise for user $u$. A sequence detector is now needed in order to equalize the channel ${\bf g}_u$. 


\noindent
{\bf Performance results with live massive MIMO channel measurements.}
We next report results for single carrier modulation in measured massive MIMO channels. Several channel measurement campaigns on massive MIMO has been conducted at Lund University, and more information about the particular one we make use of here can be found in \cite{GaEdRuTu14}. In brief, 4 users were placed outdoors around the EE-building at Lund University separated by roughly 30 meters, and a linear 128 element antenna array was placed at the roof of the building. The users were placed without any line-of-sight to the base station. The measurement bandwidth is 50 MHz and several snapshots of the propagation channel were taken. In Fig.~\ref{fig:qam_mm} we report results for the no FTN case, i.e., $\delta_{\rm t}=1$. The pulse $p(t)$ is RRC-shaped with 20\% excess bandwidth. We report averaged ASE values over the four users for the system described previously for 4-QAM and 16-QAM, using 16 or 128 antenna elements.
  The curves marked with ``AWGN'' show the results obtained when we artificially remove all intersymbol and interuser interference and therefore constitute upper bounds.    As can be seen, there is a clear gain in going from $N_{\rm BS}=16$ to $N_{\rm BS}=128$, and for 4-QAM, the gap to the upper bound is closed. With 16-QAM, the  intersymbol and interuser interference has not fully vanished, which shows up as a loss compared with the upper bound. To see how strong the interuser interference is, we also test the case of a single user, i.e., $U=1$. In this case, the gap to the bound reduces, but is not fully closed. This means that the intersymbol interference is stronger than what the single-tap equalizer used can handle. Moreover, the ASE values can be boosted by switching to FTN transmission.
  
In Fig.~\ref{fig:ftn_mm} we repeat the experiments from Fig.~\ref{fig:qam_mm}, but we use complex Gaussian modulation symbols and activate FTN; in all cases we use $N_{\rm BS}=128$ and $U=4$. In this test we assume an equalizer that can optimally deal with the intersymbol interference, but treats the interuser interference as noise. As a benchmark system, we show the ASE for the impractical but optimal sinc pulse. As we can see, there is a loss in ASE by using the RRC pulse with $\delta =0.2$. By using FTN, part of this loss is overcome as the ASE curve moves closer to the curve for the sinc pulse. Altogether, we have demonstrated that with massive MIMO, much of the intersymbol and interuser interference can vanish, so that a single-tap equalizer works well for single carrier modulation systems. With FTN activated, the loss of the excess bandwidth is reduced. With more advanced transceiver schemes, for example based on interference cancellation, the gap to the upper bounds in Fig.~\ref{fig:qam_mm} can be 
reduced, and taken together with the favorable PAPR of single carrier, this modulation format seems to be a good choice for uplinks of 5G whenever large antenna arrays can be facilitated.

Although we presented here results for the single carrier only, it is reasonable to foresee that a similar behavior can be observed also for multicarrier modulation formats, such as FBMC and OFDM: in fact in a massive MIMO system when the number of receiving antennas is sufficiently high, the interuser and intersymbol interference introduced by the channel tend to vanish, no matter the modulation adopted. Such a property has been called ``self-equalization'' and reported also in \cite{massiveMIMOFBMC}.

\section*{\small INTERACTIONS WITH 5G ARCHITECTURE AND REQUIREMENTS}

In this section we finally discuss the interactions between the reviewed modulation formats and some key requirements and features of 5G networks. Indeed, although a complete description of how a 5G cellular system will look like is not yet available, some pieces of the puzzles are already known and almost unanimously taken for granted \cite{whatwillbe}. Some of the concepts discussed here are also summarized in Table \ref{table_smiley}.

\paragraph*{Large data rates}
5G networks will have to support very large data rates; such a goal will be accomplished through a combination of technologies such as the use of multiple antennas (as already discussed), the use of adaptive modulation schemes and of course the use of larger bandwidths. This fact tends to promote the use of a multicarrier modulation for two reasons: (a) adaptive modulation is easily implemented with multicarrier schemes, wherein smart bit loading algorithms may permit to tune the modulation cardinality and the coding rate according to the channel status on each subcarrier; and (b) the use of larger bandwidths leads to increased multipath distortion, thus implying that using a multicarrier scheme simplifies the task of equalization with respect to a single carrier modulation.

\paragraph*{Small cells and mm-wave communications}
The use of small cells is a key technique aimed at increasing the overall capacity of wireless networks, intended as offered throughput per square km; recently, there has also been a growing interest for mm-wave communications \cite{6515173,pi2011introduction} for supporting short-range cellular communications. Although there is still little knowledge about mm-wave propagation in urban areas, studies are ongoing \cite{RappaportGutierrezBen-DorMurdockQiaoTamir2013}. It is anticipated that mm-wave will be used on short distances, thus implying that line-of-sight links might  be available. In this case, we will have large bandwidths, rather stable propagation environments, and low Doppler offsets. The design of a modulation scheme suited for these conditions is still an open problem, although again multicarrier schemes appear to be much more suited than single-carrier schemes.
Due to their anticipated stable propagation environments and low Doppler levels, small cell networks may be especially suitable application areas for non-orthogonal modulation formats. For FBMC, channel estimation gets inherently more challenging due to the interference at the receiver side. However, with increased stability of the propagation environment and low Dopplers, this burden gets significantly simplified.
The same arguments also apply to, e.g., advanced FBMC equalizers that equalize the interference among the symbols. Such equalizers need to be updated frequently in the case of non-negligible Doppler levels, which may impose hefty complexity increases compared with OFDM where only a single tap per detected symbol needs to be updated.
On the other hand, there is also a line of thought that foresees, for these high frequencies and large bandwidths, the use of simple modulations formats with low spectral efficiencies, deferring to future generations of cellular systems the task of optimizing the spectrum usage in these bands. The recent paper \cite{mmwaverecent},
instead, proposes the use of a single-carrier modulation with ciclic prefix as a remedy to the PAPR problem of multicarrier schemes.

\paragraph*{Uncoordinated access - internet of things}
In the coming years there will be a tremendous increase in the number of connected devices \cite{atzori2010internet,wu2011m2m}. Indeed, the current trend is to include a wireless transceiver in almost every electronic gadget/equipment, and researchers have been investigating for some years the so-called {\em internet of things} -- This is also called Machine-to-Machine (M2M) communications. A large number of connected devices will require to access the network to transmit short messages. The challenge posed by the internet of things lies, rather than in a capacity shortage, in the overwhelming burden that it produces on the signaling functions of the network. 
Regarding this aspect, the use of FBMC modulations is preferable with respect to classical OFDM since it allows uncoordinated (i.e. asynchronous) access to the subcarriers. This is one of the main messages conveyed by the ongoing 5GNOW european research project.

\paragraph*{Low latency}
Another requirement for 5G wireless cellular systems is the possibility to ensure low latency communications with a target roundtrip delay of 1ms. This is seen as a major change of 5G network with respect to existing LTE networks, since it will enable the so-called tactile Internet \cite{tactile}, which will permit the development of brand new real-time applications for monitoring and control. 
To reduce latency at the physical layer,  a single-carrier modulation seems to be preferable, since it avoids block-processing of the data which introduces additional delays. A tunable OFDM system, with an adaptive choice of the length of the data block would also be an option.

\paragraph*{Energy efficiency}
It is expected that 5G cellular networks will be far more energy-efficient than previous cellular systems \cite{tombaz2011energy}. Energy saving is mainly a matter that regards higher-layer of the network protocol stack, since it involves adaptive base station switch on/off algorithms, use of renewable energy sources, design of energy-harvesting protocols, base station sharing among network operators during off-peak hours, etc. However, at the physical layer, adaptively switching off unused carriers is a key strategy that may be used to save energy from the RF transceiver chain of base stations. This thus once again promotes the use of multicarrier systems with respect to single-carrier modulation.

\paragraph*{Cloud techniques and software radio}
Another fascinating feature of future wireless networks is the possibility of having a cloud-based radio access network \cite{zhu2011virtual,lin2010wireless}. In practice, base stations will be substituted by light devices, performing baseband-to-RF conversion and signal transmission, and connected through wired optical links to a data center, wherein data coding/decoding and higher-layer functionalities such as resource allocation will take place. The advantages of this structure are represented by the fact that centralized/cooperative strategies (such as the well-known coordinated multipoint) can be readily implemented, as well as by the fact that data modulation can be implemented by a software running in a data center. This adds a lot of flexibility to the choice of the modulation format, in the sense that paves the way to adaptive modulation schemes wherein not
only the cardinality and the coding rate may be tuned, but even the waveform itself, including the cyclic prefix; the recent 5G overview \cite{whatwillbe} thus proposes the use of "tunable OFDM," a sort of adaptive scheme with parameters chosen based on the instantaneous operating conditions.

Thus, according to the channel conditions, to the requested throughput, and to the available resources in terms, e.g., of number of antennas, adaptive schemes may be designed wherein the modulation format itself is a parameter to be optimized. We believe that, of all the key characteristics of 5G networks, the integration of cloud and software defined networking strategies within the 5G architecture will be the one to have the greatest impact on the definition of the future modulation format.

%

\section*{\small CONCLUSIONS}
This paper has provided a review of some linear modulation schemes alternative to OFDM and deemed as suitable candidates for the implementation of the air interface of future 5G cellular communications. A comparison of these modulation schemes in terms of ASE in a cellular environment has been carried out. Our results have shown that there are alternatives to OFDM offering increased values of spectral efficiency, as well as that there is no definite winner, in the sense that the preferable modulation format depends on the considered scenario in terms of channel Doppler spread, channel delay spread, and some other parameters, such as, e.g., the allowed receiver complexity. In this sense, the virtualization of the air interface and the implementation of a cloud radio access network may pave the way towards the adoption of a tunable, adaptive modulation, wherein waveform parameters are chosen based on the specific considered scenario.
The paper has also reported some discussion on the use of TFS in massive MIMO systems, and has presented a discussion on how the modulation format impacts and is impacted by key technologies and requirements of future 5G networks.

\section*{\small ACKNOWLEDGMENTS}

The authors would like to thank Professor Fredrik Tufvesson and his research group for providing the massive MIMO channel measurements.

\section*{\small REFERENCES}
\begin{small}

\bibliographystyle{IEEEbib}
\bibliography{finalRefs}

\begin{thebibliography}{10}

\bibitem{ghosh2010fundamentals}
A.~Ghosh, J.~Zhang, J.~G. Andrews, and R.~Muhamed,
\newblock {\em Fundamentals of {LTE}},
\newblock Pearson Education, 2010.

\bibitem{ochiai2001distribution}
H.~Ochiai and H.~Imai,
\newblock ``On the distribution of the peak-to-average power ratio in {OFDM}
  signals,''
\newblock {\em IEEE Trans. Commun.}, vol. 49, no. 2, pp. 282--289, Feb. 2001.

\bibitem{Morelli2004296}
M.~Morelli,
\newblock ``Timing and frequency synchronization for the uplink of an {OFDMA}
  system,''
\newblock {\em IEEE Trans. Commun.}, vol. 52, no. 2, pp. 296--306, Feb. 2004.

\bibitem{hwang2009ofdm}
T.~Hwang, C.~Yang, G.~Wu, S.~Li, and G.~Ye~Li,
\newblock ``{OFDM} and its wireless applications: a survey,''
\newblock {\em IEEE Trans.~Veh.~Tech.}, vol. 58, no. 4, pp. 1673--1694, May
  2009.

\bibitem{irmer2011coordinated}
R.~Irmer, H.~Droste, P.~Marsch, M.~Grieger, G.~Fettweis, S.~Brueck, H.-P.
  Mayer, L.~Thiele, and V.~Jungnickel,
\newblock ``Coordinated multipoint: {C}oncepts, performance, and field trial
  results,''
\newblock {\em IEEE Commun. Mag.}, vol. 49, no. 2, pp. 102--111, Feb. 2011.

\bibitem{haykin2005cognitive}
S.~Haykin,
\newblock ``Cognitive radio: {B}rain-empowered wireless communications,''
\newblock {\em IEEE J.~Select.~Areas Commun.}, vol. 23, no. 2, pp. 201--220,
  Feb. 2005.

\bibitem{Matz201387}
G.~Matz, H.~Bolcskei, and F.~Hlawatsch,
\newblock ``Time-frequency foundations of communications: {C}oncepts and
  tools,''
\newblock {\em IEEE Signal Processing Mag.}, vol. 30, no. 6, pp. 87--96, Nov.
  2013.

\bibitem{Farhang-Boroujeny201192}
B.~Farhang-Boroujeny,
\newblock ``{OFDM} versus filter bank multicarrier,''
\newblock {\em IEEE Signal Processing Mag.}, vol. 28, no. 3, pp. 92--112, May
  2011.

\bibitem{csahin2012}
A.~{\c{S}}ahin, I.~G{\"u}ven{\c{c}}, and H.~Arslan,
\newblock ``A survey on prototype filter design for filter bank based
  multicarrier communications,''
\newblock {\em IEEE Communications Surveys \& Tutorials}, Dec. 2013.

\bibitem{Matz20111}
G.~Matz and F.~Hlawatsch,
\newblock {\em Fundamentals of time-varying communication channels},
\newblock F. Hlawatsch and G. Matz, Eds., Academic Press, 2011.

\bibitem{wunder20125gnow}
G.~Wunder, M.~Kasparick, S.~ten Brink, F.~Schaich, T.~Wild, I.~Gaspar,
  E.~Ohlmer, S.~Krone, N.~Michailow, A.~Navarro, et~al.,
\newblock ``{5GNOW}: Challenging the {LTE} design paradigms of orthogonality
  and synchronicity,''
\newblock in {\em Proc.~Vehicular Tech.~Conf.}, Dresden, Germany, June 2013.

\bibitem{fusco2008sensitivity}
T.~Fusco, A.~Petrella, and M.~Tanda,
\newblock ``Sensitivity of multi-user filter-bank multicarrier systems to
  synchronization errors,''
\newblock in {\em Proc.~IEEE International Symposium on Communications, Control
  and Signal Processing}, St. Julian's, Malta, Mar. 2008, pp. 393--398.

\bibitem{Ma75c}
J.~E. Mazo,
\newblock ``Faster-than-{N}yquist signaling,''
\newblock {\em Bell System Tech.~J.}, vol. 54, pp. 1450--1462, Oct. 1975.

\bibitem{LiGe03}
A.~Liveris and C.~N. Georghiades,
\newblock ``Exploiting faster-than-{N}yquist signaling,''
\newblock {\em IEEE Trans. Commun.}, vol. 47, pp. 1502--1511, Sept. 2003.

\bibitem{RuAn05}
F.~Rusek and J.~B. Anderson,
\newblock ``The two dimensional {M}azo limit,''
\newblock in {\em Proc.~IEEE International Symposium on Information Theory},
  Adelaide, Australia, Nov. 2005, pp. 970--974.

\bibitem{Rusek2009}
F.~Rusek and J.~B. Anderson,
\newblock ``Multistream faster than {N}yquist signaling,''
\newblock {\em IEEE Trans. Commun.}, vol. 57, no. 5, pp. 1329--1340, May 2009.

\bibitem{Han20091721}
F.-M. Han and X.-D. Zhang,
\newblock ``Wireless multicarrier digital transmission via {W}eyl-{H}eisenberg
  frames over time-frequency dispersive channels,''
\newblock {\em IEEE Trans. Commun.}, vol. 57, no. 6, pp. 1721--1733, June 2009.

\bibitem{BaFeCo09b}
A.~Barbieri, D.~Fertonani, and G.~Colavolpe,
\newblock ``Time-frequency packing for linear modulations: {S}pectral
  efficiency and practical detection schemes,''
\newblock {\em IEEE Trans. Commun.}, vol. 57, pp. 2951--2959, Oct. 2009.

\bibitem{MoRuCo14}
A.~Modenini, F.~Rusek, and G.~Colavolpe,
\newblock ``Faster-than-{N}yquist signaling for next generation communication
  architectures,''
\newblock in {\em European Signal Processing Conference (EUSIPCO)}, Lisbon,
  Portugal, Sept. 2014.

\bibitem{gabor1946}
D.~Gabor,
\newblock ``Theory of communication,''
\newblock {\em IET Journal of the Institution of Electrical Engineers}, vol.
  93, no. 26, pp. 429--457, Nov. 1946.

\bibitem{Strohmer2003257}
T.~Strohmer and R.~W. Heath~Jr.,
\newblock ``Grassmannian frames with applications to coding and
  communication,''
\newblock {\em Elsevier Applied and Computational Harmonic Analysis}, vol. 14,
  no. 3, pp. 257--275, May 2003.

\bibitem{ohlmer2012model}
E.~Ohlmer, M.~Jar, and G.P. Fettweis,
\newblock ``Model and comparative analysis of reduced-complexity receiver
  designs for the {LTE}-advanced {SC-FDMA} uplink,''
\newblock {\em Elsevier Physical Communication}, vol. 8, pp. 5--21, Sept. 2012.

\bibitem{Gerstacker20131}
W.~Gerstacker, F.~Adachi, H.~Myung, and R.~Dinis,
\newblock ``Broadband single-carrier transmission techniques,''
\newblock {\em Elsevier Physical Communication}, vol. 8, pp. 1--4, Sept. 2013.

\bibitem{benvenuto2010single}
N.~Benvenuto, R.~Dinis, D.~Falconer, and S.~Tomasin,
\newblock ``Single carrier modulation with nonlinear frequency domain
  equalization: {A}n idea whose time has come again,''
\newblock {\em Proceeding of the IEEE}, vol. 98, no. 1, pp. 69--96, Jan. 2010.

\bibitem{Wang200029}
Z.~Wang and G.~B. Giannakis,
\newblock ``Wireless multicarrier communications: {W}here {F}ourier meets
  {S}hannon,''
\newblock {\em IEEE Signal Processing Mag.}, vol. 17, no. 3, pp. 29--48, May
  2000.

\bibitem{Muquet20022136}
B.~Muquet, Z.~Wang, G.~B. Giannakis, M.~De~Courville, and P.~Duhamel,
\newblock ``Cyclic prefixing or zero padding for wireless multicarrier
  transmissions?,''
\newblock {\em IEEE Trans. Commun.}, vol. 50, no. 12, pp. 2136--2148, Dec.
  2002.

\bibitem{chang1966}
R.~W. Chang,
\newblock ``Synthesis of band-limited orthogonal signals for multichannel data
  transmission,''
\newblock {\em Bell System Tech.~J.}, vol. 45, pp. 1775--1796, July 1966.

\bibitem{saltzberg1967}
B.~R. Saltzberg,
\newblock ``Performance of an efficient parallel data transmission system,''
\newblock {\em IEEE Trans.~Commun.~Technol.}, vol. 15, no. 6, pp. 805--811,
  Dec. 1967.

\bibitem{hirosaki1981}
B.~Hirosaki,
\newblock ``An orthogonally multiplexed {QAM} system using the discrete
  {F}ourier transform,''
\newblock {\em IEEE Trans. Commun.}, vol. 29, no. 7, pp. 982--989, July 1981.

\bibitem{Rugini20052279}
L.~Rugini and P.~Banelli,
\newblock ``{BER} of {OFDM} systems impaired by carrier frequency offset in
  multipath fading channels,''
\newblock {\em IEEE Trans. on Wireless Commun.}, vol. 4, no. 5, pp. 2279--2288,
  Sept. 2005.

\bibitem{Cai20032047}
X.~Cai and G.~B. Giannakis,
\newblock ``Bounding performance and suppressing intercarrier interference in
  wireless mobile {OFDM},''
\newblock {\em IEEE Trans. Commun.}, vol. 51, no. 12, pp. 2047--2056, Dec.
  2003.

\bibitem{Tomba1998580}
L.~Tomba,
\newblock ``On the effect of {W}iener phase noise in {OFDM} systems,''
\newblock {\em IEEE Trans. Commun.}, vol. 46, no. 5, pp. 580--583, May 1998.

\bibitem{Schniter20041002}
P.~Schniter,
\newblock ``Low-complexity equalization of {OFDM} in doubly selective
  channels,''
\newblock {\em IEEE Trans.~Signal Processing}, vol. 52, no. 4, pp. 1002--1011,
  Apr. 2004.

\bibitem{Mostofi2005765}
Y.~Mostofi and D.C. Cox,
\newblock ``{ICI} mitigation for pilot-aided {OFDM} mobile systems,''
\newblock {\em IEEE Trans. on Wireless Commun.}, vol. 4, no. 2, pp. 765--774,
  Mar. 2005.

\bibitem{Rugini20061}
L.~Rugini, P.~Banelli, and G.~Leus,
\newblock ``Low-complexity banded equalizers for {OFDM} systems in {D}oppler
  spread channels,''
\newblock {\em Eurasip Journal on Applied Signal Processing}, vol. 2006, pp.
  1--13, Aug. 2006.

\bibitem{Rugini2011285}
L.~Rugini, P.~Banelli, and G.~Leus,
\newblock ``{OFDM} communications over time-varying channels,''
\newblock in {\em Wireless Communications Over Rapidly Time-Varying Channels},
  pp. 285--336. F. Hlawatsch and G. Matz, Eds. Academic Press, 2011.

\bibitem{Fusco20092705}
T.~Fusco, A.~Petrella, and M.~Tanda,
\newblock ``Data-aided symbol timing and {CFO} synchronization for filter bank
  multicarrier systems,''
\newblock {\em IEEE Trans. on Wireless Commun.}, vol. 8, no. 5, pp. 2705--2715,
  May 2009.

\bibitem{Kozek19981579}
W.~Kozek and A.~F. Molisch,
\newblock ``Nonorthogonal pulseshapes for multicarrier communications in doubly
  dispersive channels,''
\newblock {\em IEEE J.~Select.~Areas Commun.}, vol. 16, no. 8, pp. 1579--1589,
  Oct. 1998.

\bibitem{RuAn06}
F.~Rusek and J.~B. Anderson,
\newblock ``On information rates of faster than {N}yquist signaling,''
\newblock in {\em Proc.~IEEE Global Telecommun. Conf.}, San {F}rancisco, CA,
  U.S.A., Nov. 2006, pp. 1--4.

\bibitem{Fang2008}
K.~Fang, L.~Rugini, and G.~Leus,
\newblock ``Low-complexity block turbo equalization for {OFDM} systems in
  time-varying channels,''
\newblock {\em IEEE Trans.~Signal Processing}, vol. 56, no. 11, pp. 5555--5566,
  Nov. 2008.

\bibitem{kay1998}
S.~M. Kay,
\newblock {\em Fundamentals of statistical signal processing, Volume 1:
  Estimation theory},
\newblock Prentice-Hall, 1998.

\bibitem{Fettweis2009}
G.~Fettweis, M.~Krondorf, and S.~Bittner,
\newblock ``{GFDM} - {G}eneralized frequency division multiplexing,''
\newblock in {\em Proc.~Vehicular Tech.~Conf.}, Barcelona, Spain, Apr. 2009,
  pp. 1--4.

\bibitem{5GNow-D3.1-2013}
M.~Kasparick, G.~Wunder, C.~F. Schaich, T.~Wild, V.~Berg, N.~Cassiau, J.~Doré,
  D.~Kténas, M.~Dryjañski, S.~Pietrzyk, I.~S. Gaspar, and N.~Michailow,
\newblock ``5{G} waveform candidate selection,''
\newblock Tech. {R}ep., D3.1 of 5G-Now, FP7 European Research Project, Nov.
  2013.

\bibitem{PHYDIAS-D3.1-2008}
J.~Louveaux, L.~Baltar, D.~Waldhauser, M.~Renfors, M.~Tanda, C.~Bader, and
  E.~Kofidis,
\newblock ``Equalization and demodulation in the receiver (single antenna),''
\newblock Tech. {R}ep., D3.1 of PHYsical layer for DYnamic AccesS and cognitive
  radio (PHYDYAS), FP7-ICT Future Networks, July 2008.

\bibitem{Ihalainen20112070}
T.~Ihalainen, A.~Ikhlef, J.~Louveaux, and M.~Renfors,
\newblock ``Channel equalization for multi-antenna {FBMC/OQAM} receivers,''
\newblock {\em IEEE Trans.~Veh.~Tech.}, vol. 60, no. 5, pp. 2070--2085, June
  2011.

\bibitem{Stamoulis20022451}
A.~Stamoulis, S.N. Diggavi, and N.~Al-Dhahir,
\newblock ``Intercarrier interference in {MIMO OFDM},''
\newblock {\em IEEE Trans.~Signal Processing}, vol. 50, no. 10, pp. 2451--2464,
  Oct. 2002.

\bibitem{Barhumi20042055}
I.~Barhumi, G.~Leus, and M.~Moonen,
\newblock ``Time-domain and frequency-domain per-tone equalization for {OFDM}
  over doubly selective channels,''
\newblock {\em Elsevier Signal Processing}, vol. 84, no. 11, pp. 2055--2066,
  Nov. 2004.

\bibitem{Das20075782}
S.~Das and P.~Schniter,
\newblock ``{Max-SINR ISI/ICI}-shaping multicarrier communication over the
  doubly dispersive channel,''
\newblock {\em IEEE Trans.~Signal Processing}, vol. 55, no. 12, pp. 5782--5795,
  Dec. 2007.

\bibitem{Caus2014}
M.~Caus and A.I. Perez-Neira,
\newblock ``Multi-stream transmission for highly frequency selective channels
  in {MIMO-FBMC/OQAM} systems,''
\newblock {\em IEEE Trans.~Signal Processing}, vol. 62, no. 4, pp. 786--796,
  Feb. 2014.

\bibitem{Caus20126519}
M.~Caus and A.I. Perez-Neira,
\newblock ``Transmitter-receiver designs for highly frequency selective
  channels in {MIMO FBMC} systems,''
\newblock {\em IEEE Trans.~Signal Processing}, vol. 60, no. 12, pp. 6519--6532,
  Dec. 2012.

\bibitem{LTE10}
{ETSI},
\newblock ``{LTE}; {E}volved {U}niversal {T}errestrial {R}adio {A}ccess
  ({E-UTRA}); {B}ase {S}tation ({BS}) radio transmission and reception (3{GPP}
  {TS} 36.104 version 11.6.0 {R}elease 11),'' Oct. 2013.

\bibitem{ZhCh03}
Y.R. Zheng and C.~Xiao,
\newblock ``Simulation models with correct statistical properties for
  {R}ayleigh fading channels,''
\newblock {\em IEEE Trans. Commun.}, vol. 51, no. 6, pp. 920--928, June 2003.

\bibitem{PHYDIAS-D5.1-2008}
Ari Viholainen, Maurice Bellanger, and Mathieu Huchard,
\newblock ``Prototype filter and structure optimization,''
\newblock Tech. {R}ep., D3.1 of PHYsical layer for DYnamic AccesS and cognitive
  radio (PHYDYAS), FP7-ICT Future Networks, Jan. 2008.

\bibitem{PiMoCoAl13}
A.~Piemontese, A.~Modenini, G.~Colavolpe, and N.~Alagha,
\newblock ``Improving the spectral efficiency of nonlinear satellite systems
  through time-frequency packing and advanced processing,''
\newblock {\em IEEE Trans. Commun.}, vol. 61, no. 8, pp. 3404--3412, Aug. 2013.

\bibitem{whatwillbe}
J.~G. Andrews, S.~Buzzi, W.~Choi, S.~Hanly, A.~Lozano, A.~C.K. Soong, and J.~C.
  Zhang,
\newblock ``What will {5G} be?,''
\newblock {\em IEEE J.~Select.~Areas Commun.}, vol. 32, no. 6, June 2014.

\bibitem{rusek2013scaling}
F.~Rusek, D.~Persson, Buon~Kiong Lau, E.G. Larsson, T.L. Marzetta, O.~Edfors,
  and F.~Tufvesson,
\newblock ``Scaling up {MIMO}: Opportunities and challenges with very large
  arrays,''
\newblock {\em IEEE Signal Processing Mag.}, vol. 30, no. 1, pp. 40--60, Jan.
  2013.

\bibitem{hoydis2013massive}
J.~Hoydis, S.~ten Brink, and M.~Debbah,
\newblock ``Massive {MIMO} in the {UL/DL} of cellular networks: How many
  antennas do we need?,''
\newblock {\em IEEE J.~Select.~Areas Commun.}, vol. 31, no. 2, pp. 160--171,
  Feb. 2013.

\bibitem{GaEdRuTu14}
X.~Gao, O.~Edfors, F.~Rusek, and F.~Tufvesson,
\newblock ``Massive {MIMO} in real propagation environments,''
\newblock {\em Available on Arxiv}, Mar. 2014.

\bibitem{massiveMIMOFBMC}
A.~Farhang, N.~Marchetti, L.~Doyle, and B.~Farhang-Boroujeny,
\newblock ``Filter bank multicarrier for massive {MIMO},''
\newblock {\em Available on Arxiv}, Feb. 2014.

\bibitem{6515173}
T.~S. Rappaport, Shu Sun, R.~Mayzus, Hang Zhao, Y.~Azar, K.~Wang, G.~N. Wong,
  J.~K. Schulz, M.~Samimi, and F.~Gutierrez,
\newblock ``Millimeter wave mobile communications for 5{G} cellular: {I}t will
  work!,''
\newblock {\em IEEE Access}, vol. 1, pp. 335--349, May 2013.

\bibitem{pi2011introduction}
Z.~Pi and F.~Khan,
\newblock ``An introduction to millimeter-wave mobile broadband systems,''
\newblock {\em IEEE Commun. Mag.}, vol. 49, no. 6, pp. 101--107, June 2011.

\bibitem{RappaportGutierrezBen-DorMurdockQiaoTamir2013}
T.~S. Rappaport, F.~Gutierrez, E.~Ben-Dor, J.~Murdock, Y.~Qiao, and J.~I.
  Tamir,
\newblock ``Broadband millimeter-wave propagation measurements and models using
  adaptive-beam antennas for outdoor urban cellular communications,''
\newblock {\em IEEE Trans.~Antennas and Prop.}, vol. 61, no. 4, pp. 1850--1859,
  Apr. 2013.

\bibitem{mmwaverecent}
A.~Ghosh, T.~A. Thomas, M.~Cudak, R.~Ratasuk, P.~Moorut, F.~W. Vook,
  T.~Rappaport, Jr. G.~R~MacCartney, S.~Sun, and S.~Nie,
\newblock ``Millimeter wave enhanced local area systems: A high data rate
  approach for future wireless networks,''
\newblock {\em IEEE J.~Select.~Areas Commun.}, vol. 32, no. 6, June 2014.

\bibitem{atzori2010internet}
L.~Atzori, A.~Iera, and G.~Morabito,
\newblock ``The internet of things: A survey,''
\newblock {\em Elsevier Computer Networks}, vol. 54, no. 15, pp. 2787--2805,
  Oct. 2010.

\bibitem{wu2011m2m}
G.~Wu, S.~Talwar, K.~Johnsson, N.~Himayat, and K.~D. Johnson,
\newblock ``{M2M}: From mobile to embedded internet,''
\newblock {\em IEEE Commun. Mag.}, vol. 49, no. 4, pp. 36--43, Apr. 2011.

\bibitem{tactile}
G.~Fettweis,
\newblock ``The tactile internet: Applications and challenges,''
\newblock {\em IEEE Vehicular Technology Magazine}, vol. 9, no. 1, pp. 64--70,
  Mar. 2014.

\bibitem{tombaz2011energy}
S.~Tombaz, A.~Vastberg, and J.~Zander,
\newblock ``Energy- and cost-efficient ultra-high-capacity wireless access,''
\newblock {\em IEEE Wireless Communications}, vol. 18, no. 5, pp. 18--24, Oct.
  2011.

\bibitem{zhu2011virtual}
Z.~Zhu, P.~Gupta, Q.~Wang, S.~Kalyanaraman, Y.~Lin, H.~Franke, and S.~Sarangi,
\newblock ``Virtual base station pool: {T}owards a wireless network cloud for
  radio access networks,''
\newblock in {\em Proc.~ACM~Intern.~Conf on Computing Frontiers}, Ischia,
  Italy, May 2011, pp. 1--10.

\bibitem{lin2010wireless}
Y.~Lin, L.~Shao, Z.~Zhu, Q.~Wang, and R.~K. Sabhikhi,
\newblock ``Wireless network cloud: Architecture and system requirements,''
\newblock {\em IBM Journal of Research and Development}, vol. 54, no. 1, pp.
  1--12, Feb. 2010.

\bibitem{ArLoVoKaZe06}
D.~M. Arnold, H.-A. Loeliger, P.~O. Vontobel, A.~Kav\v{c}i\'c, and W.~Zeng,
\newblock ``Simulation-based computation of information rates for channels with
  memory,''
\newblock {\em IEEE Trans.~Inform.~Theory}, vol. 52, no. 8, pp. 3498--3508,
  Aug. 2006.

\bibitem{RuFe12}
F.~Rusek and D.~Fertonani,
\newblock ``Bounds on the information rate of intersymbol interference channels
  based on mismatched receivers,''
\newblock {\em IEEE Trans.~Inform.~Theory}, vol. 58, no. 3, pp. 1470--1482,
  Mar. 2012.

\end{thebibliography}
\end{small}

\section*{\small AUTHORS}

\begin{small}

{\bf\em Paolo Banelli} (paolo.banelli@unipg.it) received the Laurea degree in Electronics Engineering and the Ph.D. degree in Telecommunications from the University of Perugia, Italy, in 1993 and 1998, respectively. In 2005, he was appointed Associate Professor at the Department of Electronic and Information Engineering (DIEI), University of Perugia, where he has been an Assistant Professor since 1998. In 2001, he joined as a visiting researcher, the SpinComm group, lead by Prof. G.B. Giannakis, at the Electrical and Computer Engineering Department, University of Minnesota, Minneapolis. His research interests mainly focuses  on signal processing for wireless communications, with emphasis on multicarrier transmissions, and on signal processing for biomedical applications. He is currently serving as Associate Editor of \textit{IEEE Transactions on Signal Processing}, and has been member (2011-2013) of the SPCOM Tech. Committee of the IEEE Signal Proc. Society. In 2009, he was a General Co-Chair of the IEEE 
International Symposium on Signal Processing Advances for Wireless
Communications (SPAWC). \\
{\bf\em Stefano Buzzi} (buzzi@unicas.it) is currently an Associate Professor at the University of Cassino and Lazio Meridionale, Italy. He received his  Ph.D. degree in Electronic Engineering and Computer Science from the University of Naples "Federico II" in 1999, and he has had short-term visiting appointments at the Dept. of Electrical Engineering, Princeton University, in 1999, 2000, 2001 and 2006. His research and study interest lie in the wide area of statistical signal processing and resource allocation for communications, with emphasis on wireless communications. Dr. Buzzi  is author/co-author of more than 50 journal papers and 90 conference papers; he is a former Associate Editor for the \emph{IEEE Communications Letters}, and the {\em IEEE Signal Processing Letters}, while he has recently been the lead guest editor for the  special issue on ``5G Wireless Communications Systems,'' {\em IEEE Journal on Selected Areas in Communications}, September 2014. \\
{\bf\em Giulio Colavolpe} (giulio@unipr.it) was born in Cosenza, Italy, in 1969. He received the Dr. Ing. degree in Telecommunications Engineering (cum laude) from the University of Pisa, in 1994 and the Ph.D. degree in Information Technologies from the University of Parma, Italy, in 1998. Since 1997, he has been at the University of Parma, Italy, where he is now an Associate Professor of Telecommunications at the Dipartimento di Ingegneria dell'Informazione (DII).
He received the best paper award at the 13th International Conference on Software, Telecommunications and Computer Networks (SoftCOM'05), Split, Croatia, September 2005, the best paper award for Optical Networks and Systems at the IEEE International Conference on Communcations (ICC 2008), Beijing, China, May 2008, and the best paper award at the 5th Advanced Satellite Mobile Systems Conference and 11th International
Workshop on Signal Processing for Space Communications (ASMS\&SPSC 2010), Cagliari, Italy.
He is currently serving as an Editor for \textit{IEEE Transactions on Communications} and \textit{IEEE Wireless Communications Letters}. He also served as an Editor for \textit{IEEE Transactions on Wireless Communications} and as an Executive Editor for \textit{Transactions on Emerging
Telecommunications Technologies (ETT)}. \\
{\bf\em Andrea Modenini} (modenini@tlc.unipr.it) was born in Parma, Italy, in 1986. He received the Dr. Eng. degree in telecommunications engineering (cum laude) in december 2010 from the University of Parma, Italy, where he is currently Ph.D. Student at the Dipartimento di Ingegneria dell'Informazione (DII). His main research interests include information theory and digital transmission theory, with particular emphasis on the optimization of detection algorithm from an information theoretic point of view.
He participates in several research projects funded by the European Space Agency (ESA-ESTEC) and important telecommunications companies. In the spring 2012 he was a visiting PhD student at the University of Lund, Sweden, for research on channel shortening detection for spectrally efficient modulations. \\
{\bf\em Fredrik Rusek} (fredrik.rusek@eit.lth.se)  was born in Lund, Sweden on April 11, 1978. He received the
Master of Science degree in electrical engineering in December 2002 and the Ph.D.
degree in digital communication theory in September 2007, both from Lund
Institute of Technology. In October 2007 he joined the the department of electrical
and information technology at Lund Institute and since 2012, he holds an associate professorship at the same department. Since September 2012, he is also part time employed as algorithm expert at Huawei Technologies, Lund, Sweden.
His research interests include modulation theory, equalization,
wireless communications, and applied information theory. \\
{\bf\em Alessandro Ugolini} (alessandro.ugolini@unipr.it) was born in Parma, Italy, in 1987. He received the Dr. Eng. degree in Telecommunications Engineering (cum laude) in March 2012 from the University of Parma, Italy. Since 2013 he has been a Ph.D. student at the Dipartimento di Ingegneria dell'Informazione (DII) in the same university. His main research interests include digital communications, information theory and and spectrally efficient systems. He participates in several research projects funded by the European Space Agency (ESA-ESTEC).

\end{small}


\newpage
\begin{onecolumn}

\begin{bclogo}[couleur=blue!30,arrondi=0.1, logo=\bclampe, ombre=true]{COMPUTATION OF ACHIEVABLE RATES}
\begin{small}

We sketch here a methodology for computing the ASE, that is the maximum attainable spectral efficiency with the constraint of arbitrarily small BER.
For notational simplicity, we omit the dependence of ASE on the SNR. The $\mathrm{ASE}$ takes the particular constellation and signaling parameters into consideration, so it does not qualify as a normalized capacity measure (it is often called \textit{constrained capacity}). We evaluate only ergodic rates so the $\mathrm{ASE}$ is computed given the channel realization $\mathbf{H}_{\rm c}^{\rm{(f)}}$ and averaged over it---remember that we are assuming perfect channel state information at the receiver. The spectral efficiency of any practical coded modulation system operating at a low PER is upper bounded by the $\mathrm{ASE}$, i.e., $\rho\leq \mathrm{ASE}$, where
\begin{equation}
\mathrm{ASE}=\frac{1}{T_{\rm s}F_{\mathrm{tot}}} E_{\mathbf{H}_{\rm c}^{\rm{(f)}}} \left[I(\{\mathbf{d}_{\ell}\};\{\mathbf{y}_{\ell}^{\rm{(f)}}\}|\mathbf{H}_{\rm c}^{\rm{(f)}})\right]\;\mathrm{b/s/Hz} \label{eq:ASE}
\end{equation}
$I(\{\mathbf{d}_{\ell}\};\{\mathbf{y}_{\ell}^{\rm{(f)}}\}|\mathbf{H}_{\rm c}^{\rm{(f)}})$ being the mutual information given the channel realization, and the expectation is with respect to the channel statistics.

The computation of mutual information requires the knowledge of the channel conditional probability density function (pdf) $p(\{\mathbf{y}_{\ell}^{\rm{(f)}}\}| \{\mathbf{d}_{\ell}\},\mathbf{H}_{\rm c}^{\rm{(f)}})$. In addition, only the optimal detector for the actual channel is able to achieve the ASE in (\ref{eq:ASE}). We are instead interested in the achievable performance when using suboptimal low-complexity detectors. For this reason, we resort to the framework described in~\cite[Section VI]{ArLoVoKaZe06}. We compute proper lower bounds on the mutual information
 (and thus on the ASE) obtained by substituting $p(\{\mathbf{y}_{\ell}^{\rm{(f)}}\}| \{\mathbf{d}_{\ell}\},\mathbf{H}_{\rm c}^{\rm{(f)}})$ in the mutual information definition with an arbitrary auxiliary channel law $q(\{\mathbf{y}_{\ell}^{\rm{(f)}}\}| \{\mathbf{d}_{\ell}\},\mathbf{H}_{\rm c}^{\rm{(f)}})$
with the same input and output alphabets as the original channel (mismatched detection~\cite{ArLoVoKaZe06}).\footnote{There is not strict need for $q(\{\mathbf{y}_{\ell}^{\rm{(f)}}\}| \{\mathbf{d}_{\ell}\},\mathbf{H}_{\rm c}^{\rm{(f)}})$ to be a valid conditional pdf, as it suffices that $q(\{\mathbf{y}_{\ell}^{\rm{(f)}}\}| \{\mathbf{d}_{\ell}\},\mathbf{H}_{\rm c}^{\rm{(f)}})$ is non-negative for this result to hold~\cite{RuFe12}.} 
If the auxiliary channel law can be represented/described as a finite-state channel, the
pdfs $q(\{\mathbf{y}_{\ell}^{\rm{(f)}}\}| \{\mathbf{d}_{\ell}\},\mathbf{H}_{\rm c}^{\rm{(f)}})$ and $q_{p}(\{\mathbf{y}_{\ell}^{\rm{(f)}}\}|\mathbf{H}_{\rm c}^{\rm{(f)}})=\sum_{\{\mathbf{d}_{\ell}\}}q(\{\mathbf{y}_{\ell}^{\rm{(f)}}\}\arrowvert\{\mathbf{d}_{\ell}\},\mathbf{H}_{\rm c}^{\rm{(f)}})P(\{\mathbf{d}_{\ell}\})$
can be computed, this time, by using the optimal maximum a posteriori symbol detector
for that auxiliary channel \cite{ArLoVoKaZe06}. This detector, that
is clearly suboptimal for the actual channel, has at its
input the sequence $\mathbf{y}_{\ell}^{\rm{(f)}}$ generated by simulation \emph{according
to the actual channel model} \cite{ArLoVoKaZe06}.  If we
change the adopted receiver (or, equivalently, if we change the auxiliary
channel) we obtain different lower bounds on the constrained capacity
but, in any case, these bounds are \emph{achievable} by those receivers,
according to mismatched detection theory~\cite{ArLoVoKaZe06}.
We thus say, with a slight abuse of terminology, that the computed
lower bounds are the SE values of the considered channel when those
receivers are employed.

This technique thus allows us to take reduced complexity receivers into account. In fact, it is sufficient to consider an auxiliary channel which is a simplified version of the actual channel in the sense that
only a portion of the actual channel memory and/or a limited number of impairments are present. 

In particular, in this paper we only consider auxiliary channel laws of the form
\begin{equation}
q(\{\mathbf{y}_{\ell}^{\rm{(f)}}\}| \{\mathbf{d}_{\ell}\},\mathbf{H}_{\rm c}^{\rm{(f)}})=\prod_\ell q(\mathbf{y}_{\ell}^{\rm{(f)}}| \mathbf{d}_{\ell},\mathbf{H}_{\rm c}^{\rm{(f)}})
\end{equation}
i.e., the processing is made on frequency-domain symbols independently
and it is also assumed that the receiver is based on a frequency-domain equalizer $\mathbf{G}$ and a symbol-by-symbol detector  and thus
\begin{equation}
q(\mathbf{y}_{\ell}^{\rm{(f)}}| \mathbf{d}_{\ell},\mathbf{H}_{\rm c}^{\rm{(f)}})\propto \exp\left\{-\frac{||\mathbf{G}\mathbf{y}_{\ell}^{\rm{(f)}} - \rm{diag}\left({\boldsymbol{\epsilon}_{\ell}}\right) \mathbf{d}_{\ell}||^2}{N_0}\right\}\,,
\end{equation}
where $N_0$ is the noise variance at the receiver.

The modulation formats are compared in terms of ASE without taking into account specific coding schemes, being understood that, with a properly designed channel code, the information-theoretic performance can be closely approached. 

\end{small}
\end{bclogo}

\newpage


\begin{table} [ht]
\begin{center}
\begin{tabular}{||>{\columncolor[gray]{0.8}}c|c|c|c|c|c||}
 \hline \hline
 \rowcolor{LightCyan}
& FBMC-QAM & FBMC-OQAM & SCM  & TFS-QAM & TFS-OQAM \\ 
\hline \hline
$N$ & $>1$  & $>1$ & $1$  & $>1$  &  $>1$  \\ 
\hline
$\delta_{\rm{t}}\delta_{\rm{f}}$ & $\geq 1$ & $\geq 0.5$ & $\ge 1 $& $<1$ & $<0.5$ \\ 
\hline
$\{d_{k,\ell}\}$ & $\mathrm{QAM\;symbols}$ & $\jmath^{k+\ell}a_{k,\ell}$ &
\begin{minipage}{3cm}
\centering
QAM symbols \\ with CP 
\end{minipage}
& $\mathrm{QAM\;symbols}$ & $\jmath^{k+\ell}a_{k,\ell}$   \\ 
\hline
$\{a_{k,\ell}\}$ & N.A. & \begin{minipage}{2.4cm} \centering Real-valued \\ PAM symbols \end{minipage} & QAM symbols & N.A. & \begin{minipage}{2.4cm} \centering Real-valued \\ PAM symbols \end{minipage}  \\ \hline
\hline
\end{tabular}
\caption{\label{table_par} Parameter settings for the discussed modulation formats  in view of the signal (\ref{system_model2}).}
\end{center}
\end{table}


\begin{table} [ht]
\begin{center}
\begin{tabular}{||>{\columncolor[gray]{0.8}}c|c|c|c|c|c|c||}
 \hline \hline
 \rowcolor{LightCyan}
& \begin{minipage}{2.3cm}\begin{center}Ease of hardware implementation \end{center}\end{minipage} & 
 \begin{minipage}{1.2cm}\begin{center}Low latency \end{center}\end{minipage} & 
 \begin{minipage}{1.3cm}\begin{center} Immunity to PAPR\end{center} \end{minipage} & 
 \begin{minipage}{2.3cm}\begin{center}Robustness to synch. errors\end{center} \end{minipage} & 
 \begin{minipage}{2.3cm}\begin{center}Coupling with Massive MIMO \end{center}\end{minipage} & 
 \begin{minipage}{2.3cm}\begin{center} Use with mm-Wave \end{center}\end{minipage} 
\\ 
\hline \hline
OFDM & \Large \checkmark & & & & \Large \checkmark & \Large \checkmark    \\ 
\hline
FBMC & & & &  \Large \checkmark & \Large \checkmark &  \Large \checkmark    \\ 
\hline
TFS  & & & & &  \Large \checkmark  &      \\ 
\hline
SCM  & & \Large \checkmark  &  \Large \checkmark &  & \Large \checkmark & \Large \checkmark     \\ \hline
\hline
\end{tabular}
\caption{\label{table_smiley} Suitability of considered modulation formats to 5G requirements and technologies.}
\end{center}
\end{table}


\begin{figure}
	\centering
	\includegraphics[scale=0.8]{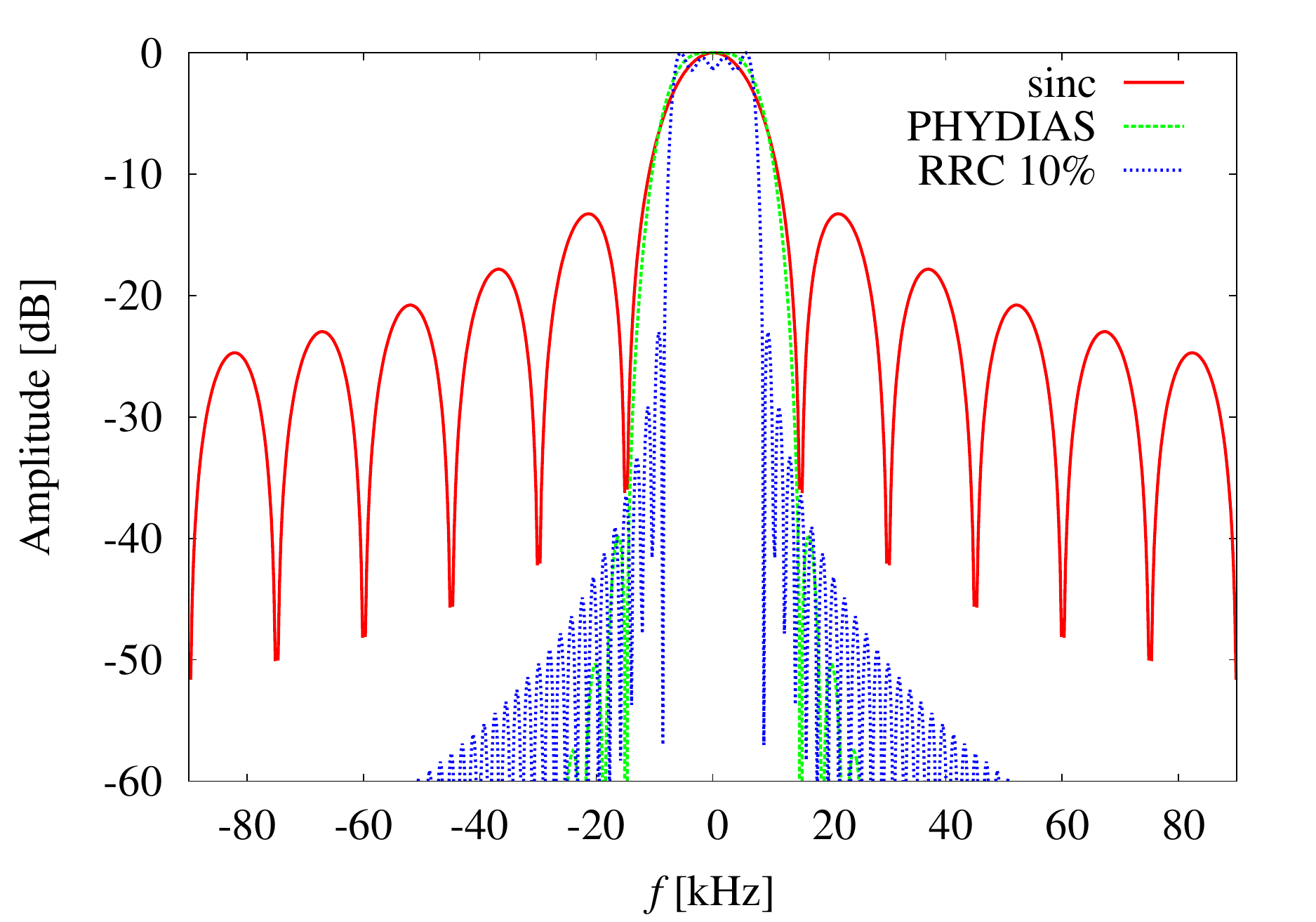}
	\caption{Spectrum of the sinc, PHYDIAS and RRC 10\% pulses.}
	\label{fig:pulses}
\end{figure}

\begin{figure}
	\centering
	\includegraphics[scale=0.8]{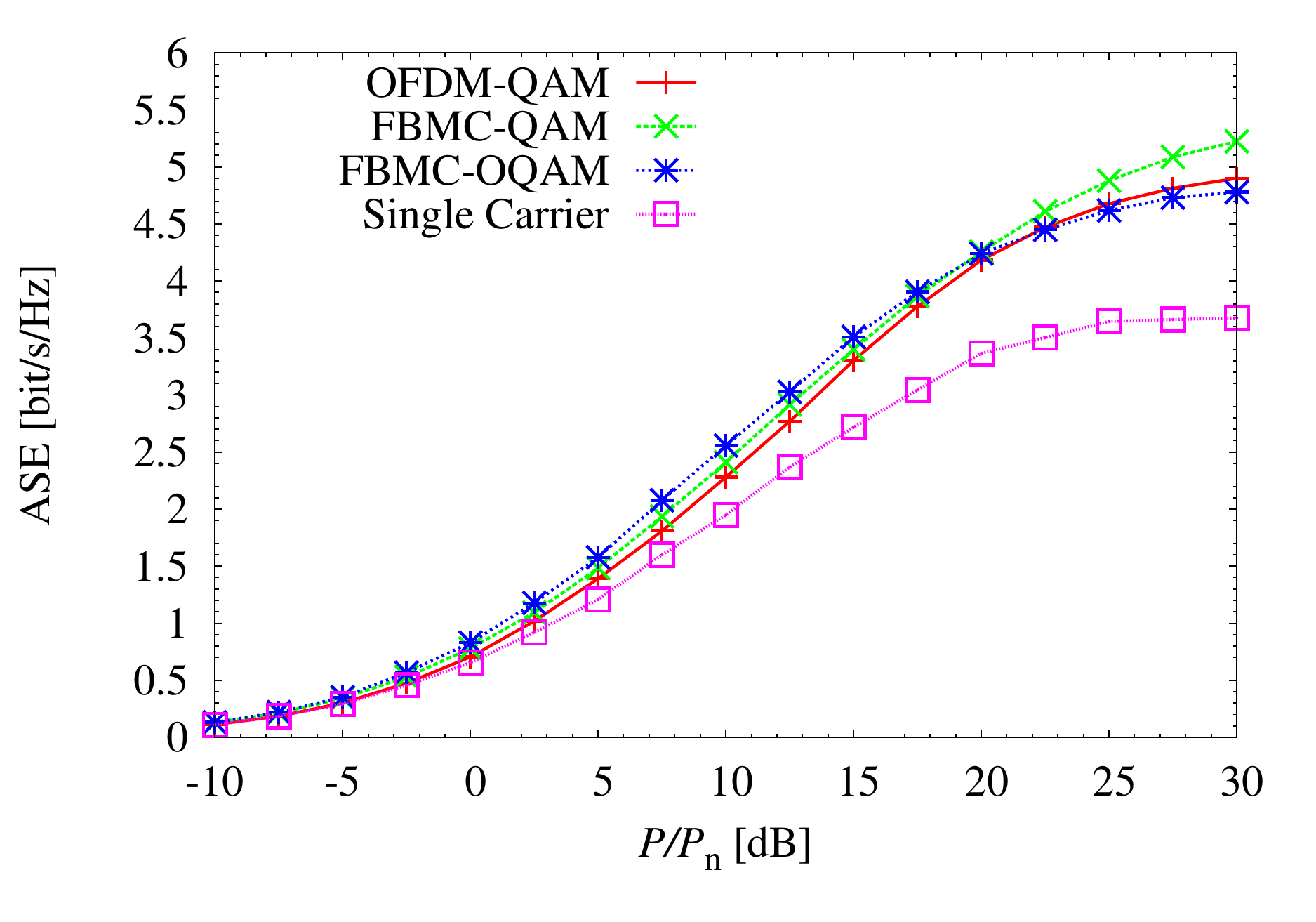}
	\caption{ASE for a low-mobility scenario: ETU channel, $f_\mathrm{d}=0$ Hz, with 64-QAM and bandwidth 1.92 MHz.}
	\label{fig:ase_etu_fd0} 
\end{figure}

\begin{figure}
	\centering
	\includegraphics[scale=0.8]{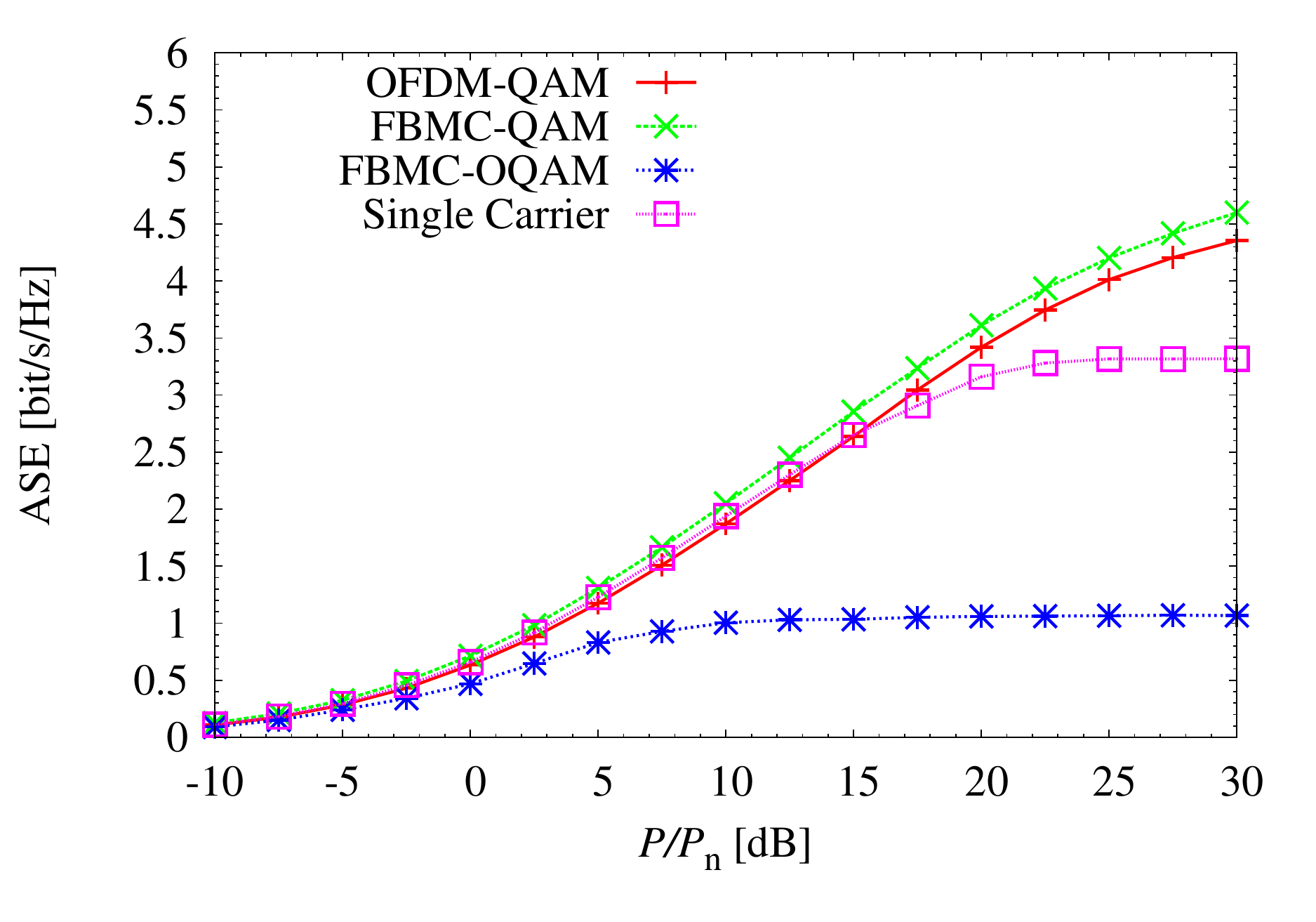}
	\caption{ASE for a high-mobility scenario: ETU channel, $f_\mathrm{d}=30$ kHz, with 64-QAM and bandwidth 1.92 MHz.}
	\label{fig:ase_etu_fd30k} 
\end{figure}

\begin{figure}
	\centering
	\includegraphics[scale=0.8]{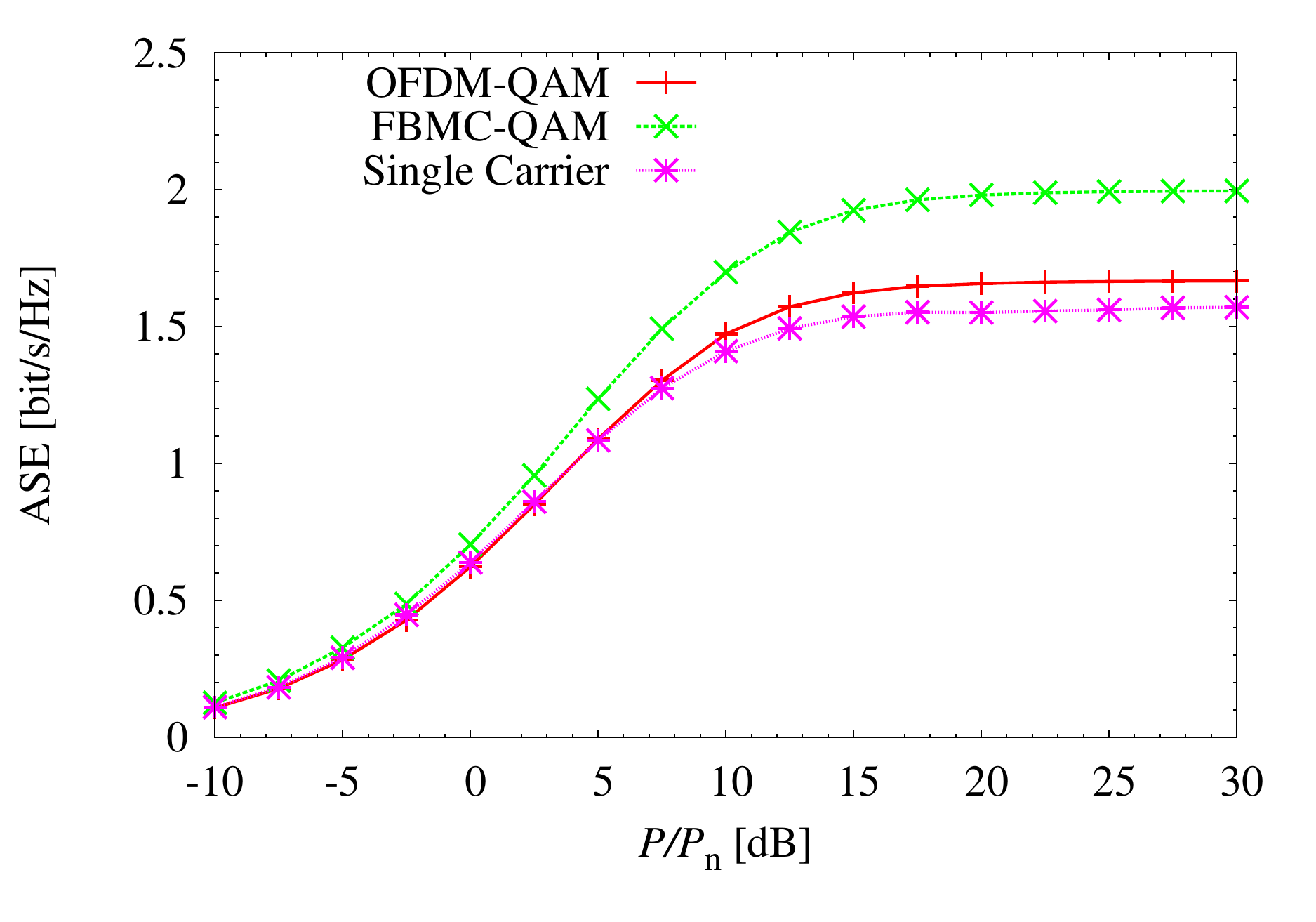}
	\caption{ASE for ETU channel, $f_\mathrm{d}=30$ kHz, with 4-QAM and bandwidth 1.92 MHz.}
	\label{fig:ase_etu_fd30k_4QAM} 
\end{figure}

\begin{figure}[htb]
	\centering
	\includegraphics[scale=0.8]{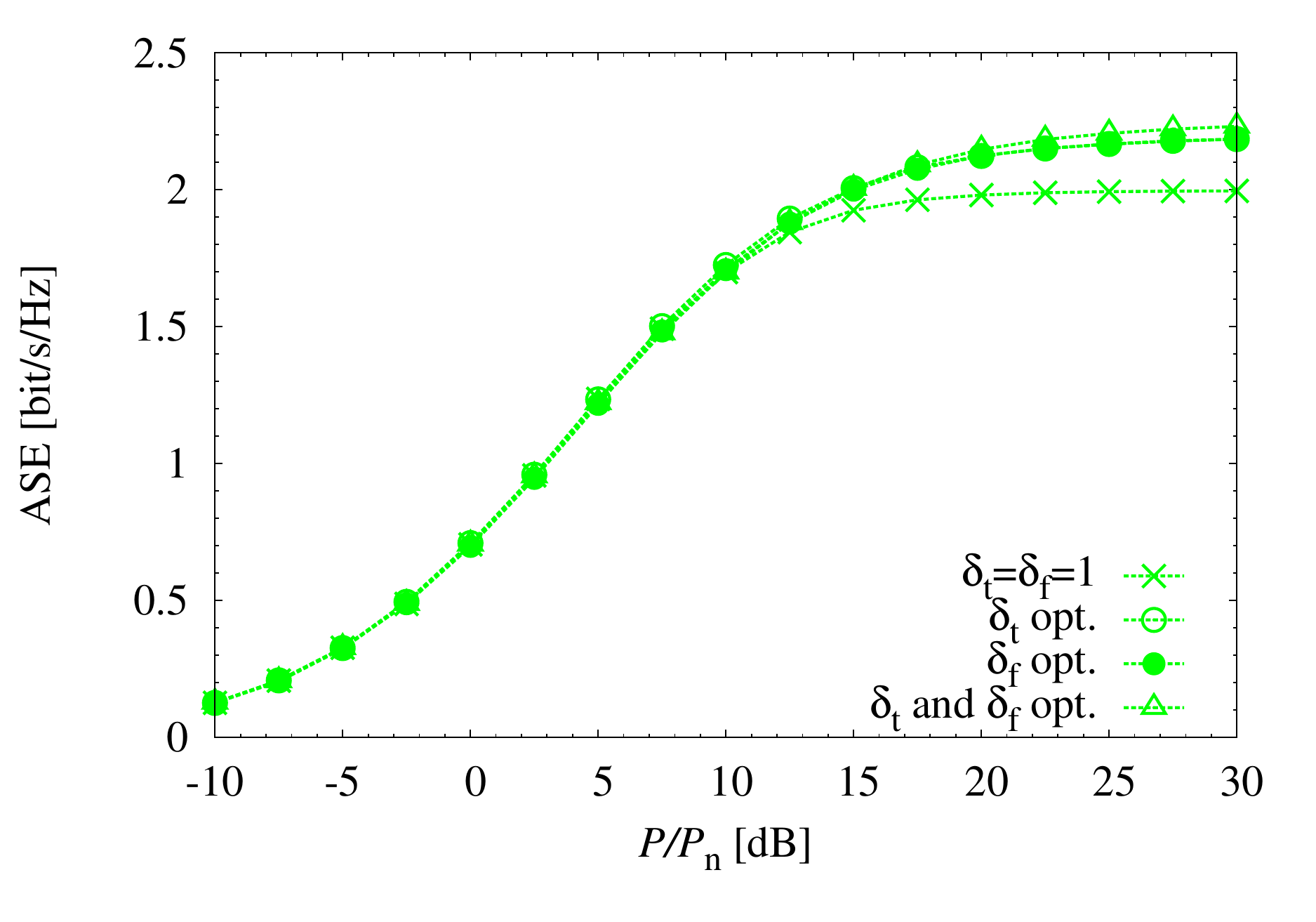}
	\caption{ASE for TFS: ETU channel, $f_\mathrm{d}=30$ kHz with 4-QAM and bandwidth 1.92 MHz.}
	\label{fig:ase_etu_tfp_f}
\end{figure}

\begin{figure}
	\centering
	\includegraphics[scale=0.8]{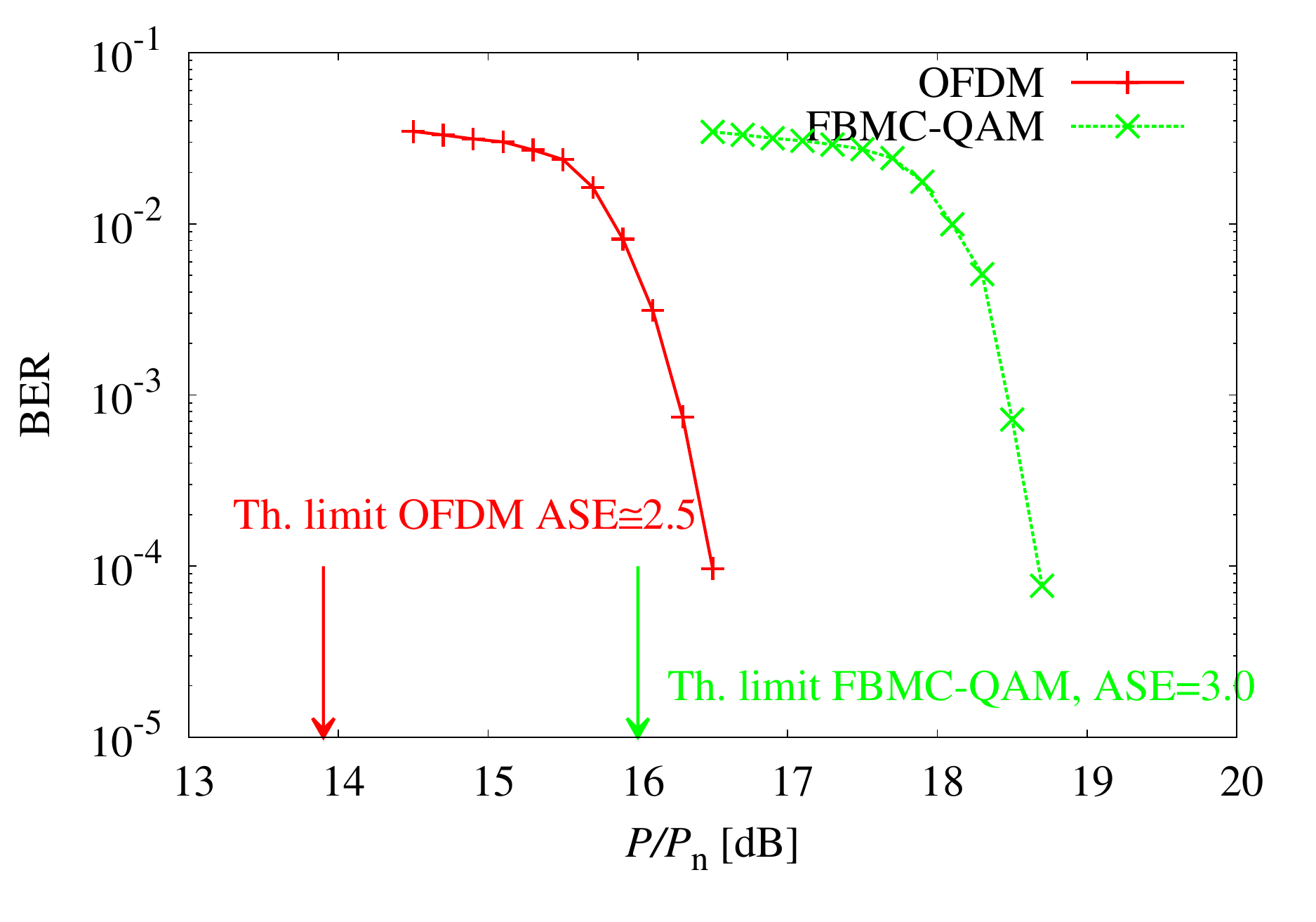}
	\caption{BER for OFDM and FBMC-QAM with 64-QAM on ETU channel, $f_\mathrm{d}=30$ kHz.}
	\label{fig:ber} 
\end{figure}

\begin{figure}
	\begin{center}
		\includegraphics[width=0.95\columnwidth]{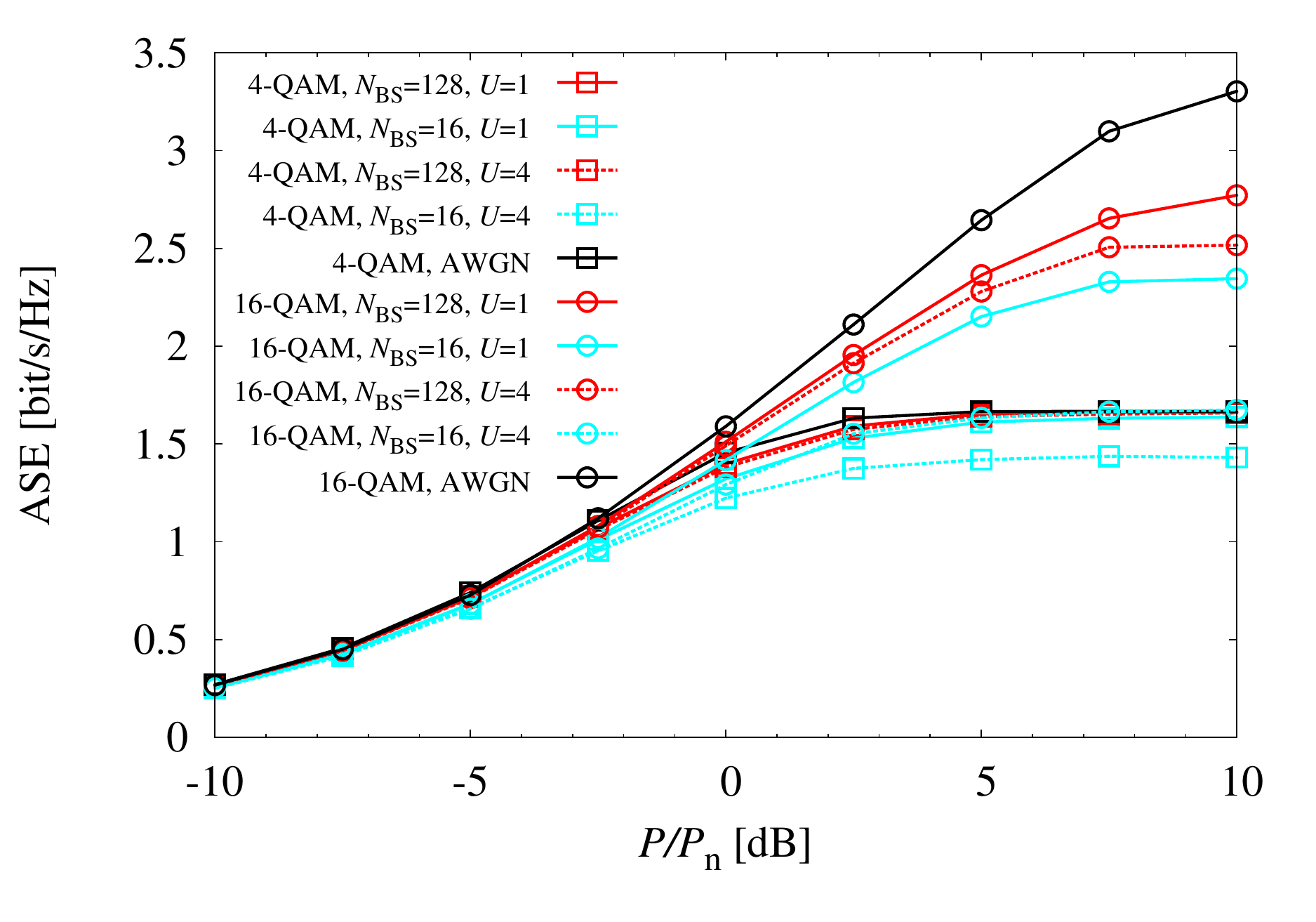}
		\caption{Averaged ASE per user for massive MIMO single-carrier FTN systems with different numbers of users and antennas, for 4-QAM and 16-QAM.}\label{fig:qam_mm}
	\end{center}
\end{figure}

\begin{figure}
	\begin{center}
		\includegraphics[width=0.95\columnwidth]{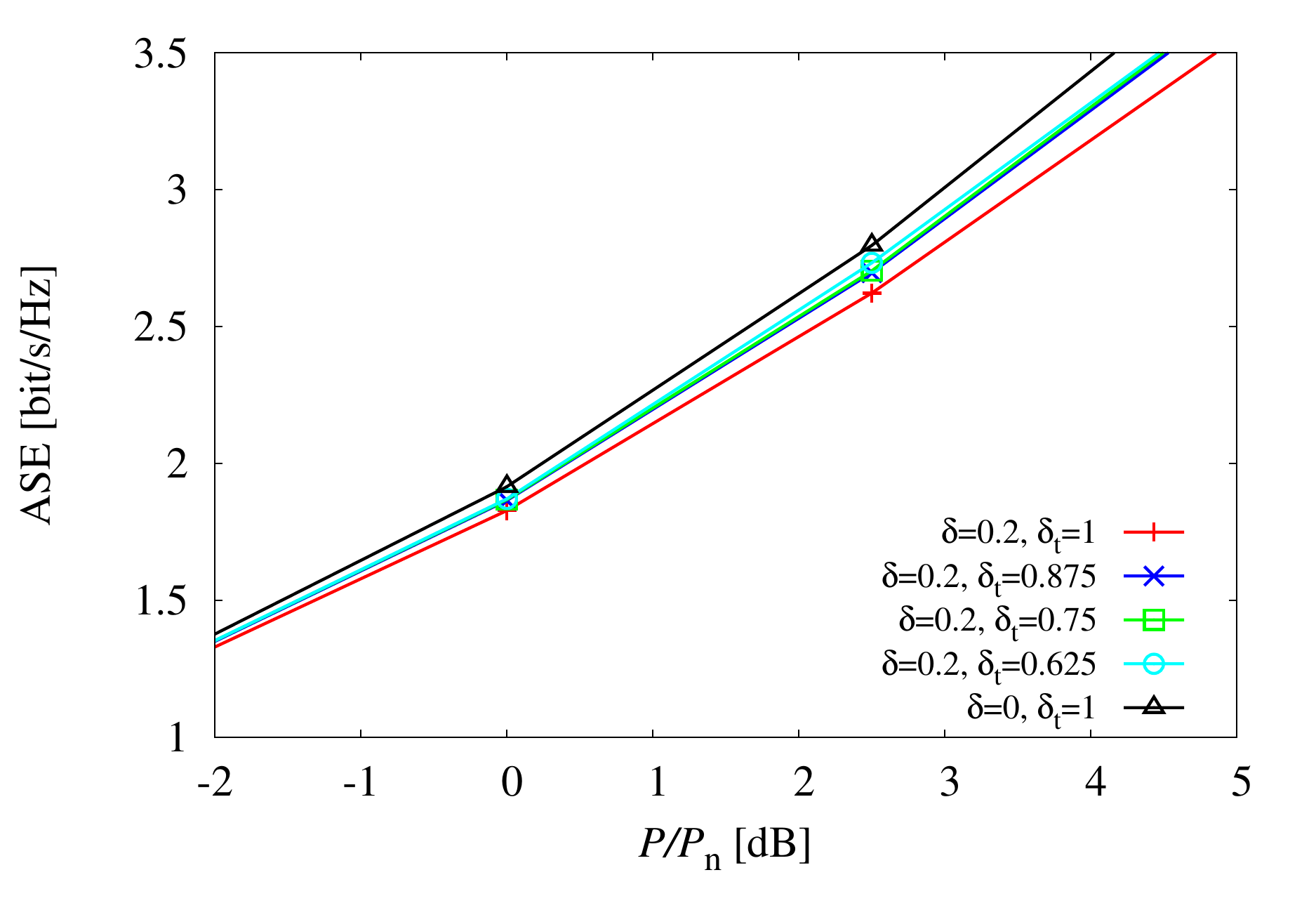}
		\caption{Averaged ASE per user for massive MIMO single-carrier FTN systems, with Gaussian inputs, $N_{\rm BS}=128$, $U=4$.}\label{fig:ftn_mm}
	\end{center}
\end{figure}

\end{onecolumn}

\end{document}